# SymPas: Symbolic Program Slicing

Yingzhou Zhang

*College of Computer, Nanjing University of Posts and Telecommunications, Nanjing 210023, China*

**Abstract:** Program slicing is a technique for simplifying programs by focusing on selected aspects of their behaviour. Current mainstream static slicing methods operate on the PDG (program dependence graph) or SDG (system dependence graph), but these friendly graph representations may be expensive and error-prone for some users. We attempt in this paper to study a light-weight approach of static program slicing, called *Symbolic Program Slicing* (SymPas), which works as a dataflow analysis on LLVM (Low-Level Virtual Machine). In our SymPas approach, slices are stored symbolically rather than procedure being re-analysed (cf. procedure summaries). Instead of re-analysing a procedure multiple times to find its slices for each callling context, SymPas calculates a single symbolic (or parameterized) slice which can be instantiated at call sites avoiding re-analysis; it is implemented in LLVM to perform slicing on its intermediate representation (IR). For comparison, we systematically adapt IFDS (Interprocedural Finite Distributive Subset) analysis and the SDG-based slicing method (SDG-IFDS) to statically IR slice programs. Evaluated on open-source and benchmark programs, our backward SymPas shows a factor-of-6 reduction in time cost and a factor-of-4 reduction in space cost, compared to backward SDG-IFDS, thus being more efficient. In addition, the result shows that after studying slices from 66 programs, ranging up to 336,800 IR instructions in size, SymPas is highly size-scalable.

## 1 INTRODUCTION AND MOTIVATION

### 1.1 Introduction

Program slicing [1] is an effective technique for narrowing the focus of attention to the relevant parts of a program. The basic idea of program slicing is to remove irrelevant statements from source codes while preserving the semantics of the program such that at some program point and/or variable, referred to as a *slicing criterion*, the variable produces the same value as its original program. A slicing criterion is a pair <$p$, $V$>, where $p$ is a program point and $V$ is a subset of program variables. Program slicing has been widely used in many software activities including software testing and debugging, measurement, re-engineering, program analysis and comprehension, and so on [2]-[7]. Program slicing can be classified into *static slicing* and *dynamic slicing* according to whether they only use static information or dynamic execution information for a specific program input. A static slice does not consider any particular execution, i.e., it works for any possible input data. Program slicing can also be classified into *backward slicing* and *forward slicing* according to the traversal direction from the slicing criterion. A backward slice consists of all statements of the program that can have some effect on the slicing criterion, whereas a forward slice contains those statements of the program that are affected by the slicing criterion. Backward slicing can assist a developer to locate the parts of the program which contain a bug. Forward slicing can be used to predict the parts of a program that will be affected by a modification. In addition, program slicing can be classified into *executable slicing* and *non-executable slicing* according to whether a slice of program $p$ is or not an executable program whose semantics are a 'subset' of the semantics of $p$ [8]. A significant portion of the non-executable code is used to support the program hierarchies. In this paper, we focus on a static non-executable and backward slicing methods, although it can be used to compute forward slices.

### 1.2 Motivation

*1.2.1 Dependency Structure Representation.* The original program slicing method was expressed as a sequence of data flow analysis problems [1], called *Weiser algorithm* for short. An alternative approach was based on reachability problem over dependence graphs such as program dependence graphs (PDG) for single-procedural programs or system dependence graphs (SDG) for multi-procedural programs [9]-[11], called as *PDG/SDG-based algorithms*, which are current mainstream slicing methods. In general, program slicing algorithms should select a data structure along with CFG (control flow graph) to store dependencies between its statements/instructions [6]. In Weiser algorithm [1], this data structure is the dataflow sets of relevant variables and relevant statements for a slicing criterion; and it is PDG/SDG in graph reachability based slicing methods. The PDG/SDG data structure makes explicit both the data and control dependences for each operation in a program [6]. **But the construction of dependence graphs such as SDG requires a larger amount of effort and suffers from blow up. Moreover, the SDG structure does not answer directly whether there is a dependency between two nodes (statements/instructions). It requires the whole SDG graph to be traversed twice for answering each dependency query.**

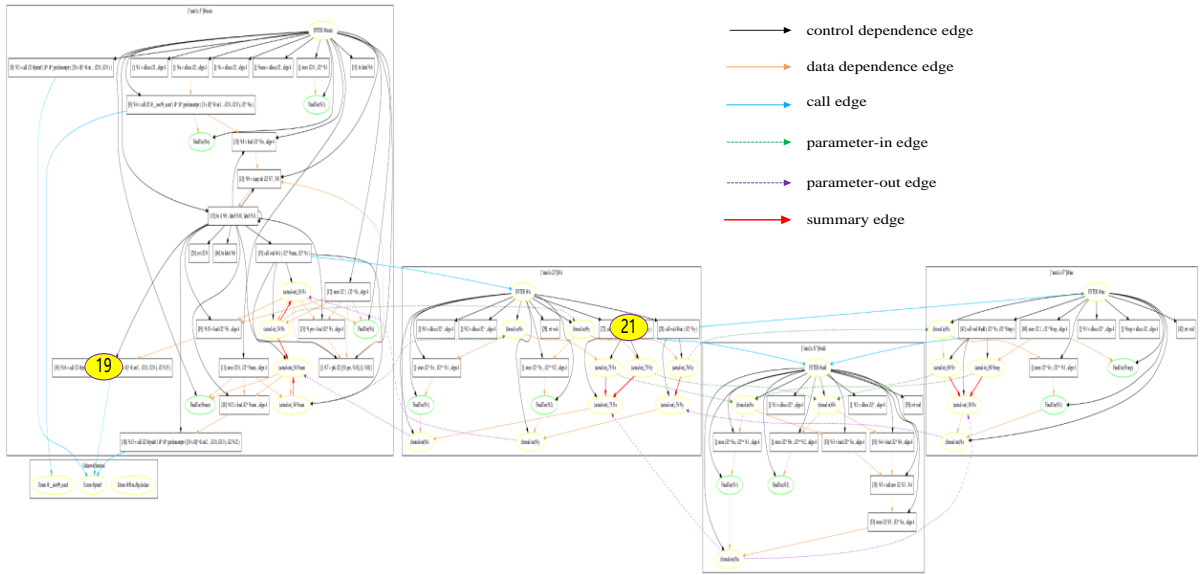

Fig. 1. **A sample SDG (system dependence graph).**

Fig. 2. **A sample instruction dependency table (IDT).**

For example, Fig. 1 is the SDG of a simple program with only 32 instructions and 4 procedures. From this SDG, if we want to query whether Instruction 19 is dependent on Instruction 21 (backward slicing), or whether Instruction 21 influences Instruction 19 (forward slicing), we need to traverse the SDG graph in two phases, instead of just checking the node reachability between Instruction 19 and 21. What's more, in each traverse phase we should carefully select different types of edges according to SDG-based algorithms [9]. These may be sophisticated and error-prone for some users. In fact, most users prefer intuitive data structures such as the instruction dependency table (IDT) in Fig. 2 (for the same program in Fig. 1), where a value of 1 indicates the presence of dependency and a value of 0 the absence of dependency. This structure of *dependency table in* Fig. 2 *can directly give the answer of the dependency query between two instructions*. For instance, since the value at Row 19 and Column 21 in Fig. 2 is 0, we know that Instruction 19 does not depend on Instruction 21, or that Instruction 21 does not influence Instruction 19. In brief, the list of values in rows corresponds to backward slice result, and the list of values in columns to forward slice result, which are very intuitive from the table. In addition, *we don't need to load the whole IDT table for each dependency query*.

Of course, IDTs (instruction dependency tables) can be computed from SDGs, but it requires large extra overhead of computations, i.e., O($n^2$), where $n$ is the number of instructions. *Can we directly choose IDT in* Fig. 2 *as the data structure along with CFG to store dependencies in slicing algorithms*? This is the target of our work.

*1.2.2 Calling-Context Problem.* SDG-based slicing algorithms have the absolute advantage over Weiser algorithm because of the calling-context problem [9]. The well-known disadvantage of the Weiser algorithm is that it cannot address the calling-context problem [9] or realizable-path reachability problem [10], e.g. the example in Fig. 3, where data flow across unmatched call/return pairs. In general, a *calling-context problem* occurs when the computation descends into a procedure $B$ that is called from a procedure $A$, it will ascend to all procedures that call $B$, not only $A$. This corresponds to execution paths which enter $B$ from $A$ and exit $B$ to a different procedure $A'$, for example, the unrealizable path $x1 \to y1 \to y2 \to z2$ or $x2 \to y1 \to y2 \to z1$ in Fig. 3. As these execution paths are infeasible, taking them into consideration results in inaccurate slices. Thus Weiser algorithm may produce less precise slices than those produced by SDG-based algorithms.

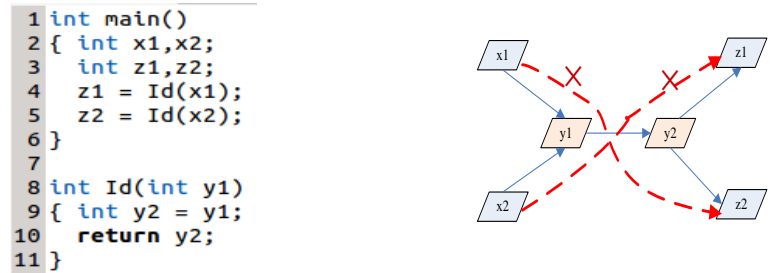

Fig. 3. **Example C program (Left) and corresponding calling-context problem (Right).**

SDG-based slicing algorithms can solve the calling-context problem using context-sensitive techniques. The fastest known SDG-based algorithm is based on the *Interprocedural Finite Distributive Subset* (IFDS) analysis [10], which is an efficient, context-sensitive and flow-sensitive dataflow analysis technique for the class of problems that satisfy its restrictions. The IFDS-based interprocedural slicing algorithm [12][10] (*RHS algorithm* for short) has been proven to be asymptotically faster than the well-known original SDG-based algorithm given by Horwitz, Reps, and Binkley [9] (*HRB algorithm* for short). Although the extended IFDS algorithm in [13] constructs the exploded supergraph on demand, these reachability algorithms require the whole exploded supergraph to be constructed ahead of time.

*Is it necessary to use these heavy-weight interprocedural analysis methods used in SDG-based algorithms to solve calling-context problem*? In fact, the precise interprocedural analysis methods such as IFDS are seem a bit expensive for some SSA-form (Static Single Assignment) [14] program languages such as LLVM IR (*Low-Level Virtual Machine*, *intermediate representation*)[15][17], where IR variables can only be assigned to once. As LLVM follows a load/store architecture, values are transferred between memory and registers via explicit load or store operations. As usual SSA form enables most of the benefits from flow-sensitive analysis to be gained from a simple flow-insensitive analysis. Therefore, *we try to explore a light-weight analysis method to address calling-context problem* in static IR slicing.

*Slice Target Language.* We select LLVM IR as the target language of program slicing, not only because it is suitable for rapid flow-sensitive analysis as mentioned above, but also because IR uses a relatively modest number of instructions (about 50) to describe a program, which is a convenient analysis target. In addition, LLVM provides an extensive infrastructure for program analysis and transformations (which preserve semantics), thus helping us to study IR slicing.

In fact, there are three reasons why we choose LLVM as our analysis infrastructure and its IR as the slice target language. 1) LLVM can help us indirectly slice multiple front-end program languages in the future. LLVM [17] is a popular and modern compiler framework that aims to support transparent program analysis and optimization for arbitrary programs. Its IR is the common code representation used throughout all phases of the LLVM compilation process. LLVM can be used as a translation of front-end source language, such as C, C++, Object C and Haskell, to LLVM IR (shown in Fig. 4) which can then be executed together in a variety of target architectures. In addition, IR slice results can be easily applied to generate the slices of front-end source languages, by extracting source codes from sliced IRs with line number information in MetaData of each IR instruction. 2) LLVM provides an extensive infrastructure for program analysis and transformations (which preserve semantics), thus allowing us to focus on just studying general IR slicing algorithms. The LLVM project provides large and continuously updated tools for IR-to-IR translations for optimization and static analysis (such as pointer analysis). Therefore, we do not need special consideration of semantic transformations before program slicing done as *conditioned program slicing* [18] and *amorphous program slicing* [19], but directly call the existing LLVM optimization tools. 3) LLVM is suitable for rapid flow-sensitive analysis. LLVM IR is in SSA [14] form, where IR variables can only be assigned to once. As LLVM follows a load/store architecture, values are transferred between memory and registers via explicit load or store operations. As usual SSA form enables most of the benefits from flow-sensitive analysis to be gained from a simple flow-insensitive analysis. In addition, LLVM IR uses a relatively modest number of instructions (about 50) to describe a program, and is fairly similar to a RISC architecture, so that it is a convenient analysis target without the details of a programming language's AST (abstract syntax tree). After source files are compiled to the LLVM IR (often in *bitcode* files), we can load them and analyze the instruction stream directly.

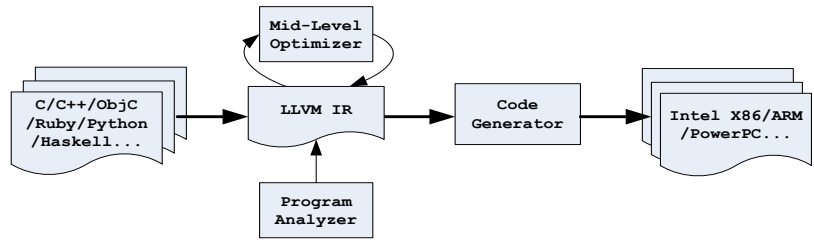

Fig. 4. **The LLVM Compiler Infrastructure.**

*1.2.3 Contributions.* In summary, in this paper, we study a **light-weight static slicing** method, called ***symbolic program slicing*** (**SymPas**), where instruction-dependency structure is the instruction dependency table (IDT), and program slices are stored symbolically rather than procedure being re-analysed (cf. procedure summaries). We use symbolic slice of procedure parameters to propagate data dependence of parameters in a similar way as parameter passing, thus helping us to solve calling-context problem.

We make the following main contributions:

- We propose a light-weight and context-sensitive static slicing approach, *symbolic program slicing* (SymPas).
- We realize the novel slicing approach SymPas for LLVM IR, which is publicly available in [20], and conduct empirical evaluations on both open-source and benchmark programs. Our results show that the SymPas approach is both efficient and size-scalable.
- For comparison, we implement two other IR slicers, **MWeiser** and **SDG-IFDS**, using *Weiser algorithm* and *SDG-based algorithm* with IFDS, respectively. To our best knowledge, this is the first adaptation of the SDG- and IFDS-based approach to statically slice IR programs. Based on these IR slicers (**MWeiser**, **SDG-IFDS** and **SymPas**), we collect their final slice results for some open-source and benchmark programs.

The rest of the paper is organized as follows: In Section 2, we introduce the data structure of IR-instruction dependency. Based on this structure, we present the backward symbolic slicing algorithm for single-procedural IR programs. In Section 3, we discuss in detail the symbolic slice for parameter dependencies and the method of backward symbolic slicing for multi-procedural IR programs. We introduce forward symbolic slicing of IR programs in Section 4. The implementation and complexity analysis of our symbolic slicing algorithm are given in Section 5, where we implement two other IR slicers by existing slicing methods to ease comparison with SymPas. In Section 6, we compare our symbolic slicing method with related slicing methods in Section 6. We in Section 7 conclude this paper with the discussion about some applications and limitations of SymPas.

## 2 INTRA-PROCEDURAL SYMBOLIC IR SLICING

### 2.1 Syntax of LLVM IR

LLVM IR represents a program as a collection of function definitions, each containing a set of basic blocks with an explicit control-flow graph (shown in Fig. 5). Each basic block is a sequence of LLVM instructions, ending in exactly one terminator instruction, which explicitly specifies its successor basic blocks [15][16]. LLVM requires that values start with a prefix: global values (functions, global variables) begin with the '@' character, and local values (register names, types) begin with the '%' character [17]. In addition, named values are represented as a string of characters with their prefix, for example, %sum and @main. Unnamed values, which are never stored, are represented as a numeric value with their prefix, for example, %10 and @3. In Fig. 5, 'phi' instruction takes a list of pairs as arguments, with one pair for each predecessor basic block of the current block. Then the non-constant variable assigned by 'phi' instruction maybe have multiple values, we call 'phi' instruction multi-valued. Similarly, 'select' instruction is also multi-valued.

In LLVM, virtual registers (local variables) are kept in SSA form, making the def-use chains explicit, i.e., each virtual register is assigned a value in exactly one instruction, and each use of a register is dominated by its definition. Memory locations (such as C arrays and structures) are kept in non-SSA form, making things very simple. Programs transfer values between registers and memory only via RISC-like load and store operations using typed pointers.

```
Module          mod    ::= layout namedtyp glb* func+ exfunc*
Global          glb    ::= id = global typ const
ExternFunction  exfunc ::= declare typ id(arg*)
Function        func   ::= define typ id(arg*) {blk+}
BasicBlock      blk    ::= l inst+
Value           val    ::= id | const
BinOp           bop    ::= add | sub | mul | fadd | fsub | fmul | udiv | sdiv | urem
                         | srem | frem | and | or | xor | shl | lshr | ashr
TruncOp         trop   ::= trunc | zext | sext | fptrunc | fpext | fptoui | bitcast
                         | fptosi | uitofp | sitofp | ptrtoint | inttoptr
Instruction
                inst   ::= br val blk_1 blk_2 | br blk | switch val blk_0 [val_j, blk_j]+ | indirectbr val blk
                         | id = cmpxchg val_1 val_2 val_3 | id = load (typ) val | id = getelementptr val const+
                         | id = icmp cond typ val_1 val_2 | id = extractvalue typ val_1 const+ | store val_1 val_2
                         | option id = call typ val param | option id = invoke typ val param | resume val
                         | id = phi typ [val_j, l_j]+ | id = alloca typ val | id = shufflevector typ val_1 val_2 val_3
                         | id = insertvalue typ val_1 val_2 const+ | id = fcmp cond typ val_1 val_2 | id = bop val_1 val_2
                         | trop val to typ | id = select val_0 typ val_1 val_2 | unreachable | id = atomicrmw val_1 val_2
                         | id = landingpad typ val padclause | id = va_arg typ val | ret val | fence ordering scope
```

Fig. 5. **Syntax of LLVM IR** [16].

( Here "*" denotes a (possibly empty) list; "+" denotes a non-empty list. For simplicity, we omit some syntax definitions such as *cond* (comparison operators) and *typ* (types). )

## 2.2 Intraprocedural Symbolic Slicing Algorithm

*2.2.1 Slicing Criterion.* In this section, we will show how to compute backward static slices of IR programs. *For simplicity, we only consider end slicing for a single variable, i.e. the slicing criterion is <p, v>, where v is the variable of interest, and p the program end point.* One can easily generalize this to a set of points and a set of variables at each point by taking the union of the individual slices [5]. As an example of IR slice, we consider an IR program in Fig. 6 (b), whose compiled C source program and CFG are shown in Fig. 6 (a) and Fig. 6 (c), respectively. A static slice of this IR program with respect to the slicing criterion <21, %a> includes all instructions except the strikethrough and shaded ones shown in Fig. 6 (b).

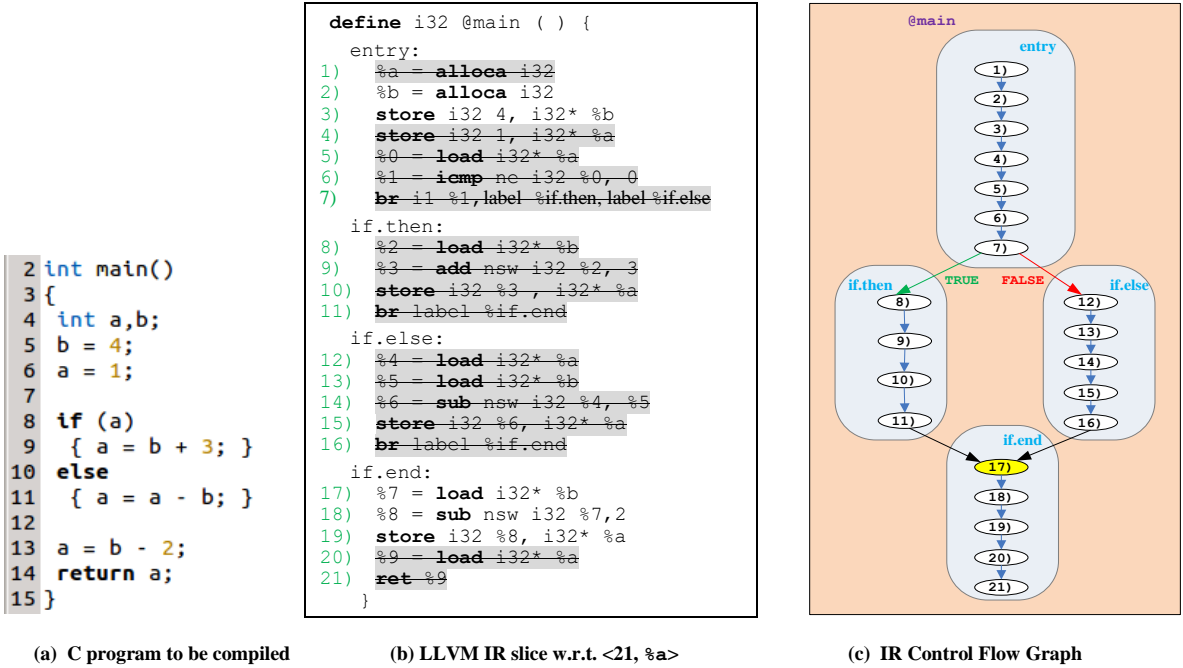

(a) C program to be compiled  (b) LLVM IR slice w.r.t. <21, %a>  (c) IR Control Flow Graph

Fig. 6. **Example of LLVM IR Slice**

*2.2.2 Instruction Dependency.* In order to get a slice from an IR program, we first compute the control and data dependences between its instructions. We choose CD($i$) to denote $i$'s control-dependence instructions; for LLVM IR, we then have **CD($i$) ≡ { $j$ | $i$ ∈ INFL($j$)}**, where INFL($j$) means the range of influence of a branch statement $j$, which can be defined as the set of statements that are control dependent on $j$ [2]. For LLVM IR, it is the range of influence from a branch instruction, $j$, such as 'switch', 'indirectbr' or 'br' in Fig. 5, which is often used to transfer control flow to one of several different basic blocks in the current function.

The IR CFG makes the control dependences of IR instructions explicit, so we will focus on data dependences. As mentioned in Section 1, related information is used to annotate CFGs at each node. In this paper, we try to use intuitive data structure, i.e. **IDT (instruction dependency table)**, to represent the dependencies between instructions. In order to avoid the sparse matrix of IDT, we introduce *instruction backward-dependence set* (**IDS**), *L($i$)*, to store dependency information for each CFG node $i$, i.e. all instructions/values that may influence $i$'s execution/computation. $L(i)$ can be obtained from Equation 1 as follows, whose initialization $L^0(i) = \{i\}$.

$$L^{k+1}(i) \equiv L^k(i) \cup \bigcup_{j \in \text{CD}(i)} L^k(j) \cup \bigcup_{x \in \text{REF}(i)} S^k(x) \tag{1}$$

where $k \geq 0$, denotes the number of computation iterations, corresponding to the level of indirect effect; REF($i$) returns the set of variables (LLVM non-constant values) referenced (used) in $i$. For example, at node 3 in Fig. 6, the reference set of its corresponding instruction (noted as $i_3$) is empty, i.e. REF($i_3$) = ∅, since we only consider rvalues (the values on the right part of an assignment) and non-constant values in REF(*i*). At node 5 (corresponding to instruction $i_5$) where REF($i_5$)={%a}, DEF($i_5$)={%0}, CD($i_5$)=∅, $S^k$(%a)= {1,4}, $L^k$($i_5$)={5}, from Equation 1 we have $L^{k+1}$($i_5$)= $L^k$($i_5$)∪$S^k$(%a)={1,4,5}. In Equation 1, $S(x)$ represents the current backward-static-slice result of *x* before analyzing *i*, whose computation equation is as follows.

$$S^{k+1}(x) \equiv \begin{cases} L^{k+1}(i) & \text{if } x \in \text{DEF}(i) \text{ and } x \text{ is single-valued} \\ S^k(x) \cup L^{k+1}(i) & \text{if } x \in \text{DEF}(i) \text{ and } x \text{ is multi-valued} \\ S^k(x) & \text{otherwise} \end{cases} \quad (2)$$

where DEF(*i*) returns the set of variables (LLVM non-constant values) defined or modified in *i*. For LLVM IR, the non-constant value may be multi-valued such as 'phi' or 'select' instruction. So for each possible value of multi-value, we extend its slice at the definition node.

*2.2.3 Procedure Symbolic Slice.* Besides the introduction of the IDT structure, another highlight of our work is that we use some ***symbolic parameters***, say $l_x$, $l_y$,..., to give the initial static slices of the corresponding formal parameters or global variables on entry to a procedure, as shown in Equation 3 below. Here the formal parameters of a procedure *P* and the global variables used in *P* are noted as FRML(*P*) and GLOB(*P*), respectively.

$$S^0(x) \equiv \begin{cases} \{l_x\} & \text{if } x \in \text{FRML}(P) \cup \text{GLOB}(P) \\ \varnothing & \text{otherwise} \end{cases} \quad (3)$$

Formally, a ***symbolic slice*** can be represented textually as a lambda abstraction '**λx.S(x)**' where x is a symbolic parameter. The symbolic slicing uses a symbolic parameter (e.g. $l_x$) for each global variable or formal parameter (say *x*) of a procedure to initialize *x*'s slice (noted as $S^0(x)$ here), i.e. $S^0(x) = \{l_x\}$, in the first slice calculation of the procedure. The symbol $l_x$, referred to as *symbolic parameter* of *x*, is just like a **placeholder** to be substituted by a dependence set. The slice result of *x*, $S(x)$, may subsume some symbolic parameters, for example $l_x$ and $l_y$, i.e. $l_x \subseteq S(x)$ and $l_y \subseteq S(x)$. We then represent this symbolic slice textually as a lambda abstraction '$\lambda l_x l_y.S(x)$'. This initialization approach with symbolic parameters is very useful to pass and obtain the dependencies of procedure parameters in Section 3. In order to facilitate interprocedural analysis in Section 3, here we introduce a table $T_P$, called ***procedure symbolic slice***, to store a copy of the final symbolic-slice table *S* of a procedure *P*, which combines symbolic slices of named values at exit (termination) instructions (such as 'ret', 'unreachable' and 'resume') of the *P* procedure. Thus we have:

$$T_P(x) \equiv S(x) \qquad \text{for each IR named value } x \text{ in } P \quad (4)$$

*2.2.4 Symbolic Slicing Algorithm.* According to Equations 1-4 above, an intraprocedural IR slice can be computed by the algorithm `IntraIRSlice` (shown in [Algorithm 1](#)). Concretely, Equation 1 corresponds to lines 9-11 in the algorithm `IntraIRSlice`, where we compute at each CFG node $L(i)$ to store instruction dependency information. Lines 13-17, 3-5 and 19-21 implement Equation 2, 3 and 4, respectively.

The key idea of the algorithm `IntraIRSlice` can be briefly stated as follows: we first uses symbolic parameters to give the initial static slices of the corresponding formal parameters or global variables on entry to a procedure (lines 3 to 5) by Equation 3. At each CFG node, we then compute instruction backward dependences as $l'$ (line 10) by Equation 1, with storing $l'$ in the node/instruction dependency table *L* (line 11). Now, $l'$ can be used to extract IR slices. Therefore, if the node analyzed is a memory-modify/definition node, we use $l'$ to update or extend the IR slice of variables (lines 13 to 17, according to Equation 2). The algorithm continues in this manner until the slice table is stable (line 6), i.e., the computation reaches its fixpoint.

A static slice includes the statements that possibly affect the variable in the slicing criterion. Therefore, for capturing these possible instructions, at each CFG join node (which has at least two predecessors), we merge the slice tables of these predecessors as the current slice table, as shown in line 8 of the algorithm. For example, at the joint node 17 in [Fig. 6](#) (c), its slice table (noted as $S_{17}$) contains the union of $S_{11}$ and $S_{16}$, i.e. $S_{17} = S_{11} \cup S_{16}$. After the algorithm finishes, the resulting slice table includes static slices for all variables (LLVM name values) of an IR program. Note that the algorithm produces another output, the IDT table *L*, which is what we need as mentioned in Section 1; and it is useful for computing forward slices (see Section 4).

Line 11 of the algorithm means that after node *i* is analyzed/computed, its IDS set $L(i)$ must be updated by the above set $l'$ (line 10), thus adding corresponding control and data dependencies. In fact, only if the instruction *i* analyzed is a branch IR instruction such as 'switch' 'indirectbr', or 'br' (excluding unconditional branch) [17], which is often used to transfer control flow to one of several different basic blocks in the current function, the set $l'$ must be added to the IDS set of each instruction of these basic blocks for propagating control dependences. Therefore, in order to improve the efficiency of [Algorithm 1](#) for backward slicing, we update $L(i)$ with the set $l'$ if *i* is a branch instruction. In other words, *for backward symbolic slicing, we* *only need to compute l' by Equation 1, at branch, value-insert or memory-modify instructions rather than at all instructions, thus saving implementation time and space.*

---

Algorithm 1. **Intraprocedural Symbolic IR-Slicing Algorithm**

*Input*: the CFG $G$ of a procedure $P$
*Output*: instruction dependency table $L$, current static slice table $S$,
and procedure symbolic slice $T_P$

```
1:   IntraIRSlice (G) {
2:   Initialize the table (L, S or T) to empty one if it does not exist
3:   for each x ∈ FRML(P)∪GLOB(P) do                    // By Equation 3
4:       if S(x) = ∅ then
5:           S(x) ← {l_x}              // Initialize slices with symbolic parameters
6:   while any changes for S do
7:       for each node i in G do
8:           If i is a CFG joint node, merge slice tables of its predecessors as current S
9:           if L(i) = ∅ then L(i) = {i}
10:          let l' = L(i) ∪ ⋃_{j∈CD(i)} L(j) ∪ ⋃_{x∈REF(i)} S(x)    // By Equation 1
11:          L(i) ← l'                  // Update dependency set IDS of i
12:          switch i do
13:              case value-insert or memory-modify instruction such as
                      'store', 'cmpxchg', 'atomicrmw', or 'insertvalue' :
14:                  for each x ∈ DEF(i) do                // By Equation 2
15:                      if x is one of multi-value in instr. such as 'phi' or 'select'.
16:                      then    S(x) ← l' ∪ S(x)       // Extend x's slice at i
17:                      else    S(x) ← l'              // Update x's slice at i
18:              default : skip                // Do nothing for other instructions
19:   for each named value x in P do                      // By Equation 4
20:       if T_P(x) = ∅ then
21:           T_P(x) ← S(x)             // Generate procedure symbolic slice
22:   return L, S and T_P
23: }
```

*2.2.5 An Example.* As an example of IR slice computation from Algorithm 1, we consider the procedure @add (shaded) in the sample IR program in Fig. 8, which is compiled (by LLVM Clang) from the simple C program shown in Fig. 7 (adapted from the example program in [9]).

Table 1. **Example IR slice computation for the procedure `@add` in Fig. 8**

| Node #i | Instruction i | REF(i) | DEF(i) | CD(i) | $L^{k+1}(i)$ | S | T |
|---|---|---|---|---|---|---|---|
| 24 | `; entry`<br>`%1 = load i32* %a` | {%a} | {%1} | ∅ | {24}∪$S^0$(%a) | $S^{k+1}$(%1)={24, $l_{\%a}$} | -- |
| 25 | `%2 = load i32* %b` | {%b} | {%2} | ∅ | {25}∪$S^0$(%b) | $S^{k+1}$(%2)={25, $l_{\%b}$} | -- |
| 26 | `%3 = add nsw`<br>`i32 %1, %2` | {%1, %2} | {%3} | ∅ | {24,25,26}∪<br>$S^0$(%a)∪$S^0$(%b) | $S^{k+1}$(%3) = {24,25,<br>26,$l_{\%a}$,$l_{\%b}$} | -- |
| 27 | `store i32 %3,`<br>`i32* %a` | {%3} | {%a} | ∅ | {24,25,26,27}<br>∪$S^0$(%a)<br>∪$S^0$(%b) | $S^{k+1}$(%a) = {24,25,<br>26,27,$l_{\%a}$,$l_{\%b}$}<br>$S^{k+1}$(%b) = {$l_{\%b}$} | -- |
| 28 | `return void` | ∅ | ∅ | ∅ | {28} | S(%a) =<br>{24,25,26,27}<br>S(%b) = ∅ | $T_{@add}$(%a)={24,25<br>,26,27,$l_{\%a}$,$l_{\%b}$}<br>$T_{@add}$(%b) = {$l_{\%b}$} |

```
 3 int main()
 4 {
 5  int n, i, sum;
 6  void A(int *x, int *y);
 7
 8  printf("Enter a positive number: ");
 9  scanf("%d", &n);
10
11  sum = 0;
12  i = 1;
13  while (i <= n)
14  {
15     A(&sum,&i);
16  }
17
18  printf ("sum = %d\n", sum);
19  printf ("i = %d\n", i);
20  return 0;
21 }

23 void A(int *x, int *y)
24 {
25    void add(int *a, int *b);
26    void inc(int *z);
27    add(x,y);
28    inc(y);
29 }
30
31 void add(int *a, int *b)
32 {
33    *a = *a + *b;
34 }
35
36 void inc(int *z)
37 {
38    void add(int *a, int *b);
39    int tmp = 1;
40    add(z,&tmp);
41 }
```

Fig. 7. **A simple C source program (adapted from [9]) to be compiled.**

```
define i32 @main ( ) {
  entry:
1) %n = alloca i32
2) %i = alloca i32
3) %sum = alloca i32
4) %1 = call i32 @printf(…)
5) %2 = call i32 @scanf(…,i32* %n)
6) store i32 0, i32* %sum
7) store i32 1, i32* %i
8) br label %while.cond
  while.cond:
9) %3 = phi i32 [[%.pre,
        %while.body],[1,%entry]]
10) %4 = load i32* %n
11) %5 = icmp sle i32 %3, %4
12) br i1 %5, label %while.body,
           label %while.end
  while.body:
13) call void @A (i32* %sum, i32* %i)
14) %.pre = load i32* %i
15) br label %while.cond
  while.end:
16) %6 = load i32* %sum
17) %7 = call i32 @printf (…,i32 %6)
18) %8 = load i32* %i
19) %9 = call i32 @printf (…,i32 %8)
20) ret i32 0
  }

define void @A(i32* %x, i32* %y)
{
  entry:
21) call void @add (i32* %x, i32* %y)
22) call void @inc(i32* %y)
23) ret void
}

define void @add(i32* %a,i32* %b)
{
  entry:
24) %1 = load i32* %a
25) %2 = load i32* %b
26) %3 = add nsw i32 %1, %2
27) store i32 %3,i32* %a
28) ret void
}

define void @inc (i32* %z)
{
  entry:
29) %tmp = alloca i32
30) store i32 1,i32* %tmp
31) call void @add(i32* %z, i32* %tmp)
32) ret void
}
```

Fig. 8. **A sample IR program.**
(Compiled from the C program in Fig. 7 by LLVM Clang using LLVM opt passes -basicaa and -gvn)

Table 1 shows the detailed data necessary to calculate the IR symbolic slice table for all single variables (LLVM name value) in @add. Working forwards from the CFG node 24, whose corresponding instruction is the entry of the procedure @add, noted as $i_{24}$, from Equations 1-3 we have

REF($i_{24}$)={%a}, DEF($i_{24}$)={%1}, CD($i_{24}$)=∅,

$L(i_{24})= L^0(i_{24}) \cup S^0(\%a)=\{24,l_{\%a}\}$,

$S(\%1)= L(i_{24}) =\{24,l_{\%a}\}$.

The data computation of node 25 is similar to that of node 24. At node 26 with $i_{26}$, since

REF($i_{26}$)= {%1,%2}, DEF($i_{26}$)={%3} and CD($i_{26}$) = ∅,

its IDS set

$L(i_{26}) = L^0(i_{26}) \cup S(\%1) \cup S(\%2) = \{24,25,26,l_{\%a},l_{\%b}\}$,

and the symbolic slice of its definition variable (non-constant IR value %3)

$S(\%3)=L(i_{26})=\{24,25,26,l_{\%a},l_{\%b}\}$.

Node 27 (corresponding to instruction $i_{27}$) is the 'store' instruction used to write value %3 to value %a, so the slice of %a needs to update with $L(i_{27})$, i.e.

$S(\%a)=L(i_{27})= \{24,25,26,27,l_{\%a},l_{\%b}\}$.

No change of the slice table $S$ occurs at node 28. After finishing the computation/analysis of the last node 28 (the exit instruction of @add), we obtain the final symbolic slice table $S$ for all named values (%a and %b) as follows:

$S(\%a) = \{24,25,26,27,l_{\%a},l_{\%b}\}$,

$S(\%b) = \{l_{\%b}\}$.

Here, if we just want to get the final static slice of @add, we simply remove all symbolic parameters in the symbolic slice table, i.e.

$S(\%a)= \{24, 25,26,27\}$,

$S(\%b) = \varnothing$.

The column $T$ in Table 1, which is just a copy of the final symbolic slice table $S$ of named values in @add, is useful for inter-procedural dataflow analysis in Section 3.

```
define i32 @main ( ) {
  entry:
1) %n = alloca i32
2) %i = alloca i32
3) %sum = alloca i32
4) %1 = call i32 @printf(…)
5) %2 = call i32 @scanf(…,i32* %n)
6) store i32 0, i32* %sum
7) store i32 1, i32* %i
8) br label %while.cond
  while.cond:
9) %3 = phi i32 [[%.pre,
        %while.body],[1,%entry]]
10) %4 = load i32* %n
11) %5 = icmp sle i32 %3, %4
12) br i1 %5, label %while.body,
        label %while.end
  while.body:
13) call void @A (i32* %sum, i32* %i)
14) %.pre = load i32* %i
15) br label %while.cond
  while.end:
16) %6 = load i32* %sum
17) %7 = call i32 @printf (…,i32 %6)
18) %8 = load i32* %i
19) %9 = call i32 @printf (…,i32 %8)
20) ret i32 0
  }
```

```
define void @A(i32* %x, i32* %y)
{
  entry:
21) call void @add (i32* %x, i32* %y)
22) call void @inc(i32* %y)
23) ret void
}

define void @add(i32* %a,i32* %b)
{
  entry:
24) %1 = load i32* %a
25) %2 = load i32* %b
26) %3 = add nsw i32 %1, %2
27) store i32 %3,i32* %a
28) ret void
}

define void @inc (i32* %z)
{
  entry:
29) %tmp = alloca i32
30) store i32 1,i32* %tmp
31) call void @add(i32* %z, i32* %tmp)
32) ret void
}
```

Fig. 9. **IR slice result w.r.t <32, %z> for the IR program in Fig. 8.**

*2.2.6 Slice Accuracy.* For the correctness of IntraIRSlice in Algorithm 1, we have the following lemmas, theorem and its corollaries, whose proofs are demonstrated in the Appendix. In fact, the terms $\bigcup_{j\in CD(i)} L^k(j)$ and $\bigcup_{x\in REF(i)} S^k(x)$ in Equation 1 can accurately capture control dependences and data dependences, respectively.

LEMMA 1. *For two instruction vertices, $w$ and $v$, in a PDG, if there is a direct (control or data dependence) edge from $w$ to $v$, noted as $w \rightarrow_{cf} v$, then we have $L^k(w) \subseteq L^{k+1}(v)$ for some $k$.*

LEMMA 2. *For two instruction vertices, $w$ and $v$, in a PDG, if $w$ can reach $v$, noted as $w \rightarrow^*_{cf} v$, then we have $L^k(w) \subseteq L(v)$ for some $k$.*

THEOREM 1. *Two PDG vertices (excluding* entry *vertices),* $w$ *and* $v$*, are reachable, i.e.,* $w \rightarrow^*_{c,f} v$*, iff (if, and only if),* $w \in L(v)$.

COROLLARY 1. *For two PDG vertices (instructions),* $w$ *and* $v$*, there is a direct edge from* $w$ *to* $v$*, i.e.* $w \rightarrow_{c,f} v$*, iff (if and only if)* $L^k(w) \subseteq L^{k+1}(v)$ *for some* $k$.

COROLLARY 2. *The accuracy of the intra-procedural symbolic slicing algorithm (Algorithm 1) is same with that of PDG-based slicing algorithms.*

Context sensitivity and pointer analysis are two of the most important issues affecting the final slice precision [7]. Context-sensitive symbolic slice algorithms, which will be discussed in Section 3, can address the calling-context problem. The introduction of a pointer will lead to aliasing problems (i.e. multiple variables accessing the same memory location), so we need some aliasing analysis (or pointer analysis) to obtain the corresponding data dependence information. Fortunately, LLVM has various optimizations, some of these aid slicing such as "-basicaa" (basic alias analysis) and "-gvn" (global value numbering), which allows us to just focus on context-sensitive IR-slicing analysis in the next section. For simplicity, we directly use two LLVM's alias analysis passes "-basicaa" and "-gvn" to implement identification; and treat both arrays and all members of a structure as a single entity. The "-basicaa" pass is an aggressive local analysis that knows many important facts such as: distinct globals, stack allocations, heap allocations, different fields of a structure, and indexes into arrays that statically differing subscripts can never alias. The "-gvn" pass performs global value numbering to eliminate fully and partially redundant instructions. It can eliminate local loads, dead loads and non-local loads (these require 'phi' node insertion).

## 3 INTER-PROCEDURAL SYMBOLIC IR SLICING

### 3.1 Motivating Example

As mentioned in Section 1, Weiser algorithm may include non-feasible paths within the program control flow, i.e. it cannot address the calling-context problem. Thus it loses precision when slicing multi-procedural programs [6]. As an example, consider again the IR program shown in Fig. 8. From *Weiser algorithm* [1], a static backward slice of this IR program with respect to the slicing criterion <32, %z> includes all the instructions of the program except the strikethrough and shaded ones shown in Fig. 9. However, it is clear that the double-line-through instructions (6 and 21) included in the slice cannot influence the slicing criterion. This less precise problem is due to the fact that *Weiser algorithm* does not keep information about the calling context when it traverses procedures up and down. This problem can be solved by interprocedural context-sensitive slicing. In this section, we will extend our intraprocedural IR slice algorithm `IntraIRSlice` in Algorithm 1 to produce context-sensitive symbolic slices for multi-procedural IR programs.

### 3.2 Procedure Dependency Analysis

*3.2.1 Procedure Call Dependence.* As shown in Section 2, `IntraIRSlice` (Algorithm 1) uses the IDT table $L$ with Equation 1 to store instruction dependencies. In this section, we will show how to extract procedure dependencies from these procedure symbolic slices computed by Algorithm 1, thus extending intraprocedural symbolic IR slicing to interprocedural one.

The dependence relations between procedures can be divided into two categories:

  1) control and data dependences caused by the procedure call and parameter passing, and

  2) data dependences due to reading and writing global variables.

A procedure call in essence transfers the program control at call site $i$ to another procedure (callee), so callee's control dependence can be computed as follows.

$$L^{k+1}(i) \equiv L^k(i) \cup \bigcup_{j \in \text{CD}(i)} L^k(j) \tag{5}$$

In general, procedure call dependences can be seen as control dependences [9], so Equation 5 corresponds to the control-dependence part in Equation 1. Next we try to capture data dependencies among procedures from parameter passing and global variable influence.

*3.2.2 Transitive Data Dependence.* Procedure calls pass some actual parameters to the formal parameters of callee. Here there is a matching process between formal and actual parameters. In general, programming language designers understand three forms of formal parameters: *in*, *out*, *in-out*, corresponding to call-by-value, call-by-result and call-by-value-result, respectively. We can consider global variables as forms of *in-out* parameters; consider the return value of a function as an *out* parameter, whose name is the same as the function name. So a function can be treated as equal to a procedure. The *out* or *in-out* parameters generally correspond to the variables modified/ defined at a call site. Fortunately, through the symbolic slice table of a procedure $P$, $T_P$, we can easily obtain as follows all possible *out* or *in-out* parameters of $P$.

$$\text{OUT}(P) \equiv \{x \mid T_P(x) \neq \{l_x\}, x \in \text{FRML}(P) \cup \text{GLOB}(P)\} \tag{6}$$

Similarly, we give the following equation to get *in* parameters of $P$:

$$\text{IN}(P) \equiv \{x \mid T_P(x) = \{l_x\}, x \in \text{FRML}(P) \cup \text{GLOB}(P)\} \tag{7}$$

In other words, from procedure symbolic slices, we can easily identify an *out* or *in-out* formal parameter (say $x$) if its symbolic slice is not equal to $\{l_x\}$, where $l_x$ is a symbolic parameter assigned to the initial slice of $x$ as shown in Equation 3; otherwise $x$ is an *in* parameter. For example, from $T_{\text{\%add}}$ shown in Table 1 (Column "$T$"), we know that the formal parameter %a is an *out* or *in-out* parameters, and %b is *in* parameter.

*Parameter Dependency.* Furthermore, **the introduction of *procedure symbolic slices*, which is the highlight of our work mentioned in Section 2, can help to obtain parameter-dependent information**, thus facilitating the calculation of the dependencies among parameters. For a parameter $x$ in a procedure $P$, its *parameter-dependent set*, noted as $\text{SUMM}_P(x)$, can be directly obtained as follows.

$$\text{SUMM}_P(x) \equiv \{y \mid l_y \in T_P(x), y \in \text{FRML}(P) \cup \text{GLOB}(P)\} \tag{8}$$

The above $\text{SUMM}_P$, which is similar to a kind of summary information in SDG-based slicing methods [9], represents transitive interprocedural data dependences. Equation 8 means that, for all symbolic parameters (introduced in Section 2) in $x$'s symbolic slice $T_P(x)$, their corresponding parameter variables are transitive data dependences of $x$. For example, since both $l_{\%a}$ and $l_{\%b}$ are in the set $T_{\text{\%add}}(\%a)$ shown in Table 1, from Equation 8, we know that the parameter %a is transitively data-dependent on both %a and %b, i.e. $\text{SUMM}_{\text{@add}}(\%a) = \{\%a, \%b\}$. In other words, *we can directly obtain parameter summary information from procedure symbolic slices, without extra interprocedural analysis methods such as IFDS*.

*Procedure Data Dependence.* In addition to the benefit of obtaining summary information directly from procedure symbolic slices, we now show that procedure symbolic slices can be used to compute data dependences among procedures, with avoiding calling-context problem. There are two cases of interprocedural data dependence at call sites:

*Case 1*: Callees' data dependence on callers via parameter passing, and
*Case 2*: Callees' influence on callers from their *out* or *in-out* parameters.

For *Case 1*, we just substitute/backfill symbolic parameters in procedure symbolic slices with corresponding parameter data dependences. In concrete, for a call site (CFG node/instruction) $i$ in a caller $P$, which calls a callee $Q$ with symbolic slice table $T_Q$, we substitute/backfill and compute as follows the temporary table $T'_{Q,i}$ from $T_Q$.

$$T'_{Q,i}(x) \equiv \begin{cases} T_Q(x) \cup L^{k+1}(i) \cup \bigcup_{y \in \text{SUMM}_Q(x)} \text{SPEC}_i^k(l_y) - \{l_y \mid y \in \text{SUMM}_Q(x)\} \\ \qquad \text{if } x \in \text{OUT}(Q) \wedge \text{SUMM}_Q(x) \neq \varnothing \\ \varnothing \qquad \text{if } x \in \text{IN}(Q) \wedge x \in \text{GLOB}(Q) \\ L^{k+1}(i) \qquad \text{if } x \in \text{IN}(Q) \wedge x \in \text{FRML}(Q) \\ T_Q(x) \cup L^{k+1}(i) \qquad \text{otherwise} \end{cases} \tag{9}$$

Where $\text{SPEC}_i(l_y)$ specifies how to pass down the corresponding data dependence at call sites to the symbolic parameter $l_y$, which can be obtained as follows from the current slice table $S$.

$$\text{SPEC}_i^k(l_y) \equiv \begin{cases} S^k(y) & \text{if } y \in \text{GLOB}(Q) \\ \bigcup_{z \in \text{REF}(x)} S^k(z) & \text{if } y \in \text{FRML}(Q), x \text{ is } y\text{'s actual parameter} \end{cases} \tag{10}$$

In Equation 9, the case "$x \in \text{OUT}(Q) \wedge \text{SUMM}_Q(x) \neq \varnothing$" means that we should substitute each symbolic parameter (say $l_y$) in $T_Q(x)$ with $\text{SPEC}_i(l_y)$, since $x$ has transitive data dependences and may be defined or modified in $Q$. If $x$ is a global variable and is not modified in $Q$, the caller $P$ will not influence $x$ in $Q$.

Now above $T'_{Q,i}$ in Equation 9 includes extra control (by Equation 5) and data dependencies because of $P$ calling once $Q$ at $i$, on the basis of $Q$'s procedural slice result $T_Q$. Therefore, we can use $T'_{Q,i}$ to update or extend the current slice table $S$ of $Q$, as a result of caller $P$ influencing callee $Q$. In summary, for a global value (say $x$) modified in $Q$. i.e. $x \in \text{OUT}(Q)$, its current symbolic slice $S(x)$ needs to be updated with $T'_{Q,i}$; otherwise, we just extend the symbolic slices of named value in $Q$, as shown in the first two cases of Equation 11.

---

**Algorithm 2. Interprocedural Symbolic IR-Slicing Algorithm**

*Input*: the CFG $G$ of a procedure $P$
*Output*: instruction dependence table $L$, current static slice table $S$,
and procedure symbolic slice $T_P$

```
 1:  InterIRSlice (G) {
 2:     Initialize the table (L, S or T) to empty one if it does not exist
 3:        ⋮      // Omit the same steps as lines 3-5 in Algorithm 1.
 4:     while any changes for S do
 5:        for each node i in G do
 6:           If i is joint node, merge slice tables S of its predecessors
 7:              ⋮   // Omit the same steps as lines 9-11 in Algorithm 1.
 8:           switch i do
 9:              case call instruction such as 'call' or 'invoke':
10:                 Let Q is the procedure called by P at i;
11:                 if Q is an external procedure such as 'llvm.memcpy' then do
12:                    Let x and v are the first two parameters of Q
13:                       S(x) ← L(i) ∪ ⋃_{y∈REF(v)} S(y)
14:                 else do
15:                    for each x ∈ FRML(Q)∪GLOB(Q) do             // By Equation 10
16:                       if x ∈ GLOB(Q) then   SPEC_i(l_x) ← S(x)
17:                       else do
18:                          Let x' is actual parameter in i related to x
19:                          SPEC_i(l_x) ← ⋃_{y∈REF(x')} S(y)
20:                    for each x ∈ T_Q do
21:                       switch x do                              // By Equation 9
22:                          case x∈OUT(Q) and SUMM_Q(x)≠∅ :
23:                             T'_{Q,i}(x) ← T_Q(x) ∪ L(i) ∪ ⋃_{y∈SUMM_Q(x)} SPEC_i(l_y) − {l_y | y∈ SUMM_Q(x)}
24:                          case x∈IN(Q) and x∈GLOB(Q) :   T'_{Q,i}(x) ← ∅
25:                          case x∈IN(Q) and x∈FRML(Q) :   T'_{Q,i}(x) ← L(i)
26:                          default :   T'_{Q,i}(x) ← T_Q(x) ∪ L(i)
27:                    if x∈GLOB(Q) and x∈OUT(Q)                   // By Equation 11
28:                       then   S(x) ← T'_{Q,i}(x)
29:                       else   S(x) ← S(x) ∪ T'_{Q,i}(x)
30:                    for each actual parameter x in i do
31:                       Let y is Q's formal parameter related to x
32:                       if y∈OUT(Q) then   S(x) ← T'_{Q,i}(x)
33:              ⋮    // Omit the same steps as lines 13-21 in Algorithm 1.
34:     return L, S and T_P
35:  }
```

For *Case 2* above, we need to consider the influence from the callee $Q$ to the caller $P$ at their call sites, through those parameters that may be modified in $Q$. With above $T'_{Q,i}$, the third case (in red border and shadow) of Equation 11 can capture these callees' influences on callers.

$$S^{k+1}(x) \equiv \begin{cases} T'_{Q,i}(x) & \text{if } x \in \text{GLOB}(Q) \land x \in \text{OUT}(Q) \\ S^k(x) \cup T'_{Q,i}(x) & \text{if } x \notin \text{GLOB}(Q) \land x \in T_Q \\ T'_{Q,i}(y) & \text{if } x \text{ is actual parameter in } i, \\ & \quad y \text{ is } x\text{'s formal parameter, and } y \in \text{OUT}(Q) \\ S^k(x) & \text{otherwise} \end{cases} \quad (11)$$

Table 2. **Symbolic IR slice computation for the procedure `@inc` and `@A` in Fig. 8**

| Proc. | Node #$i$ | Instruction $i$ | REF($i$) | DEF($i$) | CD($i$) | $L^{k+1}(i)$ | $S$ | $T$ |
|---|---|---|---|---|---|---|---|---|
| @inc | 29 | ; entry %tmp = **alloca** i32 | ∅ | {%tmp} | ∅ | {29} | $S^k$(%tmp)={29} | -- |
| | 30 | **store** i32 1, i32* %tmp | ∅ | {%tmp} | ∅ | {30} | $S^{k+1}$(%tmp)={30} | -- |
| | 31 | **call void** @add (i32* %z, i32* %tmp) | {%z, %tmp} | {%z} | ∅ | {31} | $S^{k+1}$(%a)={24,25,26,27, 30,31,$l_{\%z}$} $S^{k+1}$(%b)={31} $S^{k+1}$(%z)={24,25,26,27, 30,31,$l_{\%z}$} | $T'_{@add,31}$(%a)= {24,25,26,27,30,31,$l_{\%z}$} $T'_{@add,31}$(%b)={31} |
| | 32 | **return void** | ∅ | ∅ | ∅ | {32} | $S$(%a)={24,25,26,27, 30,31} $S$(%b)={31} $S$(%tmp)={30} $S$(%z)={24,25,26,27, 30,31} | $T_{@inc}$(%tmp)={30} $T_{@inc}$(%z)={24,25,26, 27,30,31,$l_{\%z}$} |
| @A | 21 | ; entry **call void** @add (i32* %x, i32* %y) | {%x, %y} | {%x} | ∅ | {21} | $S^k$(%a)={24,25,26,27,30,31, 21,$l_{\%x},l_{\%y}$} $S^k$(%b)={31,21} $S^k$(%x)={24,25,26,27, 21,$l_{\%x},l_{\%y}$} | $T'_{@add,21}$(%a)= {24,25,26,27,21, $l_{\%x},l_{\%y}$} $T'_{@add,21}$(%b)={21} |
| | 22 | **call void** @inc (i32* %y) | {%y} | {%y} | ∅ | {22} | $S^{k+1}$(%a)={24,25,26,27,30,3 1, 21,$l_{\%x},l_{\%y}$,22} $S^{k+1}$(%b)={31,21,22} $S^{k+1}$(%tmp)={30,22} $S^{k+1}$(%z)={24,25,26,27, 30,31,22,$l_{\%y}$} $S^{k+1}$(%y)={24,25,26,27,30,3 1, 22,$l_{\%y}$} | $T'_{@inc,22}$(%z)= {24,25,26,27,30,31,22,$l_{\%y}$} |
| | 23 | **return void** | ∅ | ∅ | ∅ | {23} | $S$(%a)={21,22,24,25, 26,27,30,31} $S$(%b)={21,22,31} $S$(%tmp)={22,30} $S$(%z)={22,24,25,26, 27,30,31} $S$(%x)={21,24,25,26, 27} $S$(%y)={22,24,25,26, 27,30,31} | $T_{@A}$(%x)={21,24,25,26, 27,$l_{\%x},l_{\%y}$} $T_{@A}$(%y)={22,24,25,26, 27,30,31,$l_{\%y}$} |

As described above, because of symbolic parameters in procedure symbolic slices, data dependences of parameters among procedures can be propagated in time in similar way as parameter passing. ***This timely propagation of data dependence and the strict correspondence of symbolic parameters, can effectively guarantee data flow across realizable paths, thus avoiding the calling-context problem.***

### 3.3 Interprocedural Symbolic Slicing Algorithm

*3.3.1 Context-Sensitive IR Slicing Algorithm.* According to the above interprocedural dependency analysis (Equations 9-11) with procedure symbolic slice, we present a context-sensitive IR slicing algorithm (InterIRSlice) in Algorithm 2, where call nodes/instructions can be handled with the help of procedure symbolic slices of callees, and other instructions can be analyzed in the same way as the intraprocedural IR slicing method in Algorithm 1.

Lines 15-19 in Algorithm 2 generate specific dependencies SPEC$_i$ (by Equation 10) at call instruction $i$ for all symbolic parameters in symbolic slices of callees. Lines 21-26 and 27-32 implement Equation 9 and 11, respectively. In addition, lines 11 to 13 handle the external call procedure such as '*llvm.memcpy*' as a definition instruction. Here, all callee procedures must have been analyzed before using Algorithm 2 to analyze their caller procedures. So the procedures must be analyzed in a particular order, for example, strongly connected components (SCC) in procedural call graph. The analysis starts at the leaves and propagates summary information up the call graph. For recursive procedures, SCCs are analyzed until a fixed-point is reached. Each SCC is processed as a unit, so we can analyze SCCs in parallel, thus Algorithm 2 supporting parallelised implementation.

Table 3. **Symbolic slice table *T* and final slice table *S* of the IR program in Fig. 8**

| Proc. | Var. | Symbolic slice *T* | SUMM & SPEC | Final slice result *S* |
|---|---|---|---|---|
| @add | %a | $T_{@add}(\%a) = \{24,25,26,27,l_{\%a}, l_{\%b}\}$ | SUMM$_{@add}$(%a)={%a,%b}<br>SPEC$_{31}$($l_{\%a}$) = $S^k$(%z)<br>SPEC$_{21}$($l_{\%a}$) = $S^k$(%x) | $S$(%a)={5,**6**,7,9,10,11,12,13,14,**21**,22,24,25,26,27,30,31} |
| | %b | $T_{@add}(\%b) = \{l_{\%b}\}$ | SUMM$_{@add}$(%b)={%b}<br>SPEC$_{31}$($l_{\%b}$) = $S^k$(%tmp)<br>SPEC$_{21}$($l_{\%b}$) = $S^k$(%y) | $S$(%b)={5,7,9,10,11,12,13,14,**21**,22,24,25,26,27,30,31} |
| @inc | %z | $T_{@inc}(\%z) = \{24,25,26,27,30,31,l_{\%z}\}$ | SUMM$_{@inc}$(%z)={%z}<br>SPEC$_{22}$($l_{\%z}$) = $S^k$(%y) | $S$(%z) = {5,7,9,10,11,12,13,14,22,24,25,26,27,30,31} |
| | %tmp | $T_{@inc}$(%tmp)={30} | -- | $S$(%tmp) = {5,7,9,10,11,12,13,14,22,24,25,26,27,30,31} |
| @A | %x | $T_{@A}(\%x) = \{21,24,25,26,27,l_{\%x}, l_{\%y}\}$ | SUMM$_{@A}$(%x)={%x,%y}<br>SPEC$_{13}$($l_{\%x}$) = $S^k$(%sum) | $S$(%x)={5,**6**,7,9,10,11,12,13,14,**21**,22,24,25,26,27,30,31} |
| | %y | $T_{@A}(\%y) = \{22,24,25,26,27,30,31,l_{\%y}\}$ | SUMM$_{@A}$(%y)={%y}<br>SPEC$_{13}$($l_{\%y}$) = $S^k$(%i) | $S$(%y)={5,7,9,10,11,12,13,14,22,24,25,26,27,30,31} |
| @main | %n | $T_{@main}$(%n) = {5} | -- | $S$(%n) = {5} |
| | %i | $T_{@main}$(%i) ={5,7,9,10,11,12,13,14,22,24,25,26,27,30,31} | -- | $S$(%i) = {5,7,9,10,11,12,13,14,22,24,25,26,27,30,31} |
| | %sum | $T_{@main}$(%sum)={5,6,7,9,10,11,12,13,14,21,22,24,25,26, 27,30,31} | -- | $S$(%sum)={5,**6**,7,9,10,11,12,13,14,**21**,22,24,25,26,27,30,31} |

*3.3.2 An Example.* As an example, we consider again the symbolic slicing of the sample IR program shown in Fig. 8. Based on its procedure call graph, we choose an analysis order (one of the topological sorts) of procedures, i.e. @add → @inc → @A → @main. Since there is no call instruction in @add procedure, we can easily compute its symbolic slice table, $T_{@add}$, by the IntraIRSlice algorithm as shown in Table 1 in Section 2. From $T_{@add}$, the dependency information among the formal parameters of @add can be obtained by Equations 6-8 as follows:

OUT(@add)={%a},    IN(@add)= {%b},
SUMM$_{@add}$(%a) = {%a, %b},    SUMM$_{@add}$(%b) = {%b}.

According to above analysis order, we next analyze the @inc procedure, which has a call node (the 31st instruction in Fig. 8, noted as $i_{31}$) to callee @add. Since neither %a or %b in @add is a global variable, from line 19 in Algorithm 2 or Equation 10, we have:

SPEC$_{31}$($l_{\%a}$)=$S$(%z)={$l_{\%z}$} ,
SPEC$_{31}$($l_{\%b}$)=$S$(%tmp)={30}.

Through lines 21 to 26 in the InterIRSlice algorithm, the temporary slice table, $T'_{@add,31}$, including interprocedural dependencies need to be computed as follows.

$T'_{@add,31}(\%a) = L(i_{31}) \cup (S(\%z) \cup S(\%tmp)) \cup T_{@add}(\%a) − \{l_{\%a}, l_{\%b}\}$
$= \{31\} \cup (\{l_{\%z}\} \cup \{30\}) \cup \{24,25,26,27,l_{\%a},l_{\%b}\} − \{l_{\%a},l_{\%b}\}$
$= \{24,25,26,27,30,31,l_{\%z}\}$

$$T'_{@add,31}(\%b) = L(i_{31}) = \{31\}$$

Because of @inc calling @add at $i_{31}$, we extend the symbolic slices of named values (i.e. %a and %b) in @add by using $T'_{@add,31}$ above, as line 29 in Algorithm 2. In addition, we use $T'_{@add,31}$ to update the symbolic slices of possible modified values (i.e. %z) at $i_{31}$, as line 32 in Algorithm 2. The current symbolic slice result after analysing $i_{31}$ is shown in Table 2 (Row "31" and Column "*S*").

The algorithm continues in above manner. After analysing all procedures, we can obtain procedure-symbolic-slice table *T* and current symbolic slice table *S* as shown in Table 2, with the final static slice result shown in Table 3. For example, the final slice result of %z in Table 3, $S(\%z)$, means that a static backward slice with respect to the slicing criterion <32, %z> includes all the instructions of the IR program except the double-line-through, strikethrough and shaded ones shown in Fig. 9. *In fact, its final IDT table L has already been shown in Fig. 2 in Section 1.*

### 3.3.3 Slice Accuracy.
As shown above, in our algorithm InterIRSlice, using procedure symbolic slices helps to obtain accurate parameter-dependent information, so as to facilitate the calculation of the dependencies among parameters (Equation 8) and the dependencies among procedures (Equation 9 and 11), thus avoiding the calling-context problem. About the correctness of InterIRSlice, based on Theorem 1, Lemmas 1 to 2, and Equations 5, 8 to 10, we have the following lemmas and theorem, whose proofs are shown in the Appendix. In fact, **Equation 5, Equation 10, Equation 11 and Equation 8 respectively correspond to *call* (*control*) edges, *parameter-in* edges, *parameter-out* edges and *summary* edges in a SDG** [9], which are four addition kinds of edges for connecting PDGs.

LEMMA 3. In a SDG, for a *call-site* vertex *u* in procedure *Q*, which calls procedure *P*, let *w* be an *actual-in* vertex at *u*, with corresponding to formal parameter *x* of *P*; let *v* be an *actual-out* vertex at *u*, with corresponding to formal parameter *y* of *P*. Then, there is a *summary* edge from *w* to *v*, noted as $w \rightarrow_{su} v$, iff $x \in \text{SUMM}_P(y)$.

LEMMA 4. For two SDG vertices, *w* and *v*, if there is a direct edge from *w* to *v*, i.e., $w \rightarrow v$, then we have $L^k(w) \subseteq L^{k+1}(v)$ for some *k*.

LEMMA 5. For two instruction vertices, *w* and *v*, in a SDG, if *w* can reach *v* in procedure *Q* by following *control* edges, *flow* edges and *summary* edges, noted as $w \rightarrow^*_{c,f,su} v$, then $w \in L(v)$ and $L^k(w) \subseteq L(v)$ for some *k*.

THEOREM 2. **For two instruction nodes, *w* and *v*, in a SDG *G*, let *G/v* be the backward slice of *v* via SDG-based slicing algorithms [9]. Then, $w \in G/v$ iff $w \in L(v)$.**

In short, our procedure symbolic slices with symbolic parameters, can be used to directly answer some queries such as: whether a parameter variable is modified/defined in a procedure, and whether two parameters have transitive data dependence. In addition, without extra interprocedural analysis techniques, procedure symbolic slices can also be used to obtain as follows, the set of parameter variables which can be referenced (GREF) or modified (GMOD) [9] by a procedure.

$$\text{GMOD}(P) = \{x \mid T_P(x) \neq \{l_x\}, x \in \text{FRML}(P) \cup \text{GLOB}(P)\}$$
$$\text{GREF}(P) = \{y \mid (x, S_x) \in T_P, l_y \in S_x, y \in \text{FRML}(P) \cup \text{GLOB}(P)\}$$

In [9], GMOD and GREF are additionally computed to eliminate *actual-out* and *formal-out* vertices for parameters that will never be modified, resulting in more precise slices. For example,

$$\text{GMOD}(@add) = \{\%a\}, \quad \text{GREF}(@add) = \{\%a, \%b\}.$$

## 4 FORWARD SYMBOLIC IR SLICING

For certain cases, we are interested in all those statements that may be influenced by the slicing criterion, i.e., *forward slicing*, which can be used to analyze modification propagation. For example, the forward slice with respect to <1, %n> for the sample IR program shown in Fig. 8 contains all instructions except those labeled with 1-4 and 6-8. For simplicity, just as for backward slicing, we here only consider forward slicing for a single variable, i.e. the slicing criterion is <*p*, *v*>, where *v* is the variable of interest, and *p* the program start point.

The forward slicing algorithm based on our symbolic slicing method, ForwardIRSlice, is shown in Algorithm 2. In fact, the IDS set of node *i*, $L(i)$, obtained from the InterIRSlice algorithm (Algorithm 2), includes all instructions/values that may influence *i*'s execution. On the other hand, *i* must be included in forward slices of all variables (LLVM name values) referenced in $L(i)$. This is achieved in lines 7 to 8 of the ForwardIRSlice algorithm. As shown in previous two sections, the IDT table *L* includes control and data dependencies between procedures, thus Algorithm 3 can be used to compute forward IR slices for interprocedural procedures.

---

Algorithm 3.  **Forward Symbolic Slicing Algorithm**
*Input*: the CFG $G$ of a procedure
*Output*: forward static slice table $S'$

```
1:  ForwardIRSlice (G){
2:      Initialize the table S' to the empty table
3:      // Get instruction dependence table L by calling Algorithm 2
4:      let (L,S,T) = InterIRSlice(G)
5:      for each i ∈ L do
6:          for each j ∈ L(i) do
7:              for each named value x ∈ REF(j) do
8:                  S'(x) ← {i} ∪ S'(x)
9:      return S'
10: }
```

---

## 5 IMPLEMENTATIONS

### 5.1 Implementation of Symbolic Slicing Approach

Based on our symbolic IR-slicing algorithms (Algorithm 1 to Algorithm 3) in previous sections, we implemented a symbolic program slicer based on LLVM, noted as **SymPas**, in Haskell [23] using the library *llvm-analysis* [24] for analyzing LLVM bitcodes. Our tool **SymPas** has been made available on its website [20]. All experiments were performed on a desktop PC with 4 processors (2.5GHz Intel i5), 8GB of memory and 64-bit Ubuntu GNU/Linux 12.04.

*5.1.1 Data Structures in SymPas.* In practice, in order to obtain good performance of set operations on tables such as the IDT table $L$ and the slice table $S$ in our algorithms, we choose data structures such as Haskell types `IntSet`, `IntMap` and `Map` [25], i.e. $L$ :: `IntMap IntSet` and $S$ :: `Map String IntSet`.

The data structures `IntSet` and `IntMap` are based on big-Endian Patricia trees [26] and use some bit-level coding tricks [27], so that they perform well on binary operations like union and intersection. Many of their operations (e.g. *lookup* and *insert*) have a worst-case complexity of $O(\min(m, W))$, where $m$ is the number of elements; $W$ is the number of bits in an `Int` (32 or 64). The `Map` data structure is based on size-balanced binary trees [28]; its operations such as *lookup* and *insert* have a worse-case complexity of $O(\log k)$, where $k$ is the number of its keys. What's more, *S can be embedded in L at those instructions (such as 'alloca' in Fig. 5) where slice-criterion variables are initially defined*, since the IDS (instruction backward-dependence sets) of these instructions only contain their own.

In addition, as an implementation trick, we set each symbolic parameter (e.g. $l_x$) in the symbolic slice table as a negative number of its unique ID. For example, the symbolic backward slice of `%a` at exit site of `@add` in Table 1, $T_{@add}(\%a)$ or $\{24,25,26,27,l_{\%a},l_{\%b}\}$, can be represented/stored as $\{-100,-101,24,25,26,27\}$, here assuming that the unique ID numbers of `%a` and `%b` are 100 and 101, respectively.

*5.1.2 Complexity of SymPas.* A bound on the cost of our symbolic slicing can be expressed according to the data structure of tables $L$ and $S$, which are effective maps as mentioned above. For the `IntraIRSlice` algorithm (Algorithm 1), its complexity analysis is restricted to the intermediate set $l'$ (line 10 of Algorithm 1) by Equation 1. In the case of IR program, the terms $\bigcup_{j \in CD(i)} L^k(j)$ and $\bigcup_{x \in REF(i)} S^k(x)$ in Equation 1 will respectively cost $O(b \times n \times \min(n, W))$ and $O(r \times n \times \log v)$ time, so $l'$ can be determined in $O(b \times n \times \min(n, W) + r \times n \times \log v)$ or $O(b \times n + r \times n \times \log v)$ time, where $b$ is the number of branch instructions such as 'switch','indirectbr' or 'br', $r$ the largest number of (named/unnamed) variables referred in any IR instruction, $v$ the number of single variables in the program, and $n$ the number of nodes (vertices) in the CFG, which is almost equal to the number of instructions in the IR program. Thus the intraprocedural symbolic slicing algorithm can determine a static slice in $O((b + r \times \log v) \times n \times e)$ time, where $e$ is the number of edges in the CFG. If $S$ is embedded in $L$, the time cost of `IntraIRSlice` is $O((b + r) \times n \times e)$. Below we assume that $S$ is embedded in $L$.

For the `InterIRSlice` algorithm (Algorithm 2), at procedural call instructions, besides the cost of $l'$, **SymPas** needs additional time cost of computing non-*in* parameter dependencies for backfilling the symbolic parameters in the symbolic slice table, i.e., SPEC$_i$ (lines 15-19 of Algorithm 2) by Equation 10. Its time cost is $O(r \times n \times x)$, where $x = globals + params$, *globals* is the number of global variables in the program, and *params* the largest number of formal parameters in any procedure. So each call instruction will cost additional time $O(r \times n \times x^3)$ by Equation 9, and the worst-case running time of `InterIRSlice` can be bounded by

$$O((b_{max} + r) \times p \times n_{max} \times e_{max} + r \times n_{max} \times c \times x^3),$$

where $c$ is the number of call instructions, $p$ is the number of procedures in the program, and $n_{max}$ and $e_{max}$ are the largest number of nodes and edges in any procedure's CFG, respectively.

For forward symbolic slicing in the `ForwardIRSlice` algorithm (Algorithm 3), we need extra time cost $O(r \times p^2 \times n_{max}^2)$ on Algorithm 2 to extract forward slice table.

To analyze the space complexity of the algorithms, we focus on to the construction tables $L$ and $T$. We need space $O(p^2 \times n_{max}^2)$ and $O(p \times v_{max} \times n_{max})$ to hold the IDT table $L$ and the procedure- symbolic-slice table $T$, respectively. Therefore, the total space cost is $O(p^2 \times n_{max}^2)$.

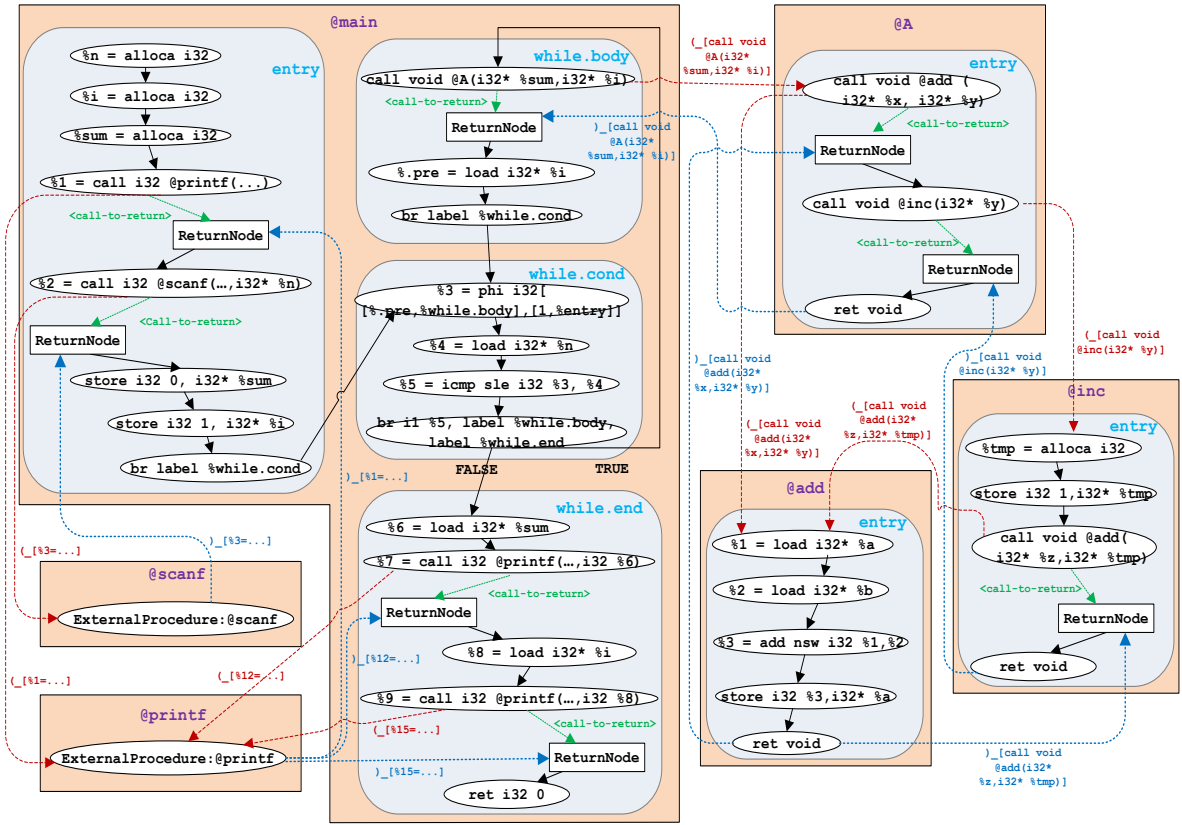

Fig. 10. **ICFG (Interprocedural CFG) for IR program in Fig. 8.**

## 5.2 IR Slicers with Existing Slicing Methods

In order to facilitate comparison of our symbolic IR slicing (SymPas) with existing slicing methods in same running environments such as implementation language, compilation configuration and optimization, we implement two other static slicers for LLVM IR: **MWeiser** (using Weiser slicing algorithm [1]) and **SDG-IFDS** (using SDG-based slicing algorithm [9] with IFDS [10][13] analysis), which are also available in [20].

The IFDS algorithm [10] finds the meet-over-all-valid-paths solution (instead of the meet-over-all-paths solution), through transforming an interprocedural dataflow-analysis problem into a special kind of graph-reachability problem (reachability along interprocedurally realizable paths). The algorithm takes the interprocedural control flow graph (ICFG, also called supergraph in [10], e.g. Fig. 10) and transforms it into an exploded supergraph. In order to explicitly track calling-context, the IFDS algorithm uses **call strings** [29], which are built by labeling the *call* and *return* for call-site $c_i$ with unique terminal symbols (e.g., "($_i$" and ")$_i$" ). The call-strings method can ensure that the string of symbols generated by a graph traversal belongs to some context-free language, e.g., each "($_i$" must be matched with ")$_i$". For example, in the ICFG (shown in Fig. 10) of the sample IR program in Fig. 8, the call edges (the red dashed line) and the return edges (the blue dotted line) are labelled by the unique symbols that begins with "(_" and ")_" strings, respectively.

Naeem et al in [13] gave some practical extensions to the IFDS algorithm for larger programs and suitable for programs in SSA form [14]. The input to the extended algorithm is a dataflow function that, given an ICFG (supergraph) node $n$, computes all of the edges leaving $n$. In [13], this dataflow function is split into four separate flow transfer functions. Based on this extended IFDS algorithm, we can compute static IR slices by explicitly defining these flow functions. So we implement the IR slicer **SDG-IFDS** by providing its corresponding flow transfer functions in LLVM.

As mentioned in Section 1, program slicing algorithms should select a data structure along with CFGs to store dependencies between statements/instructions. In data-flow equations based slicing methods such as *Weiser algorithm*, this data structure is the data-flow set such as def-use chains; It is PDG/SDG or ICFG supergraph, in graph reachability based slicing methods such as *HRB algorithm* and *RHS algorithm*; In our symbolic slicing method (SymPas), we choose the intuitive instruction-dependence table (IDT) as the data structure with CFG for storing instruction dependencies. Most of the existing slicing algorithms rely on relation graphs such as PDG/SDG.

In Table 4, we compare our symbolic slicing method (SymPas) with three existing slicing methods mentioned above: *Weiser*, *HRB* (SDG-based) and *RHS* (SDG-IFDS-based) algorithms. As mentioned before (esp. in Section 3), SymPas is a context-sensitive slicing method via forward dataflow analysis on LLVM. It uses intuitive data structure IDT to represent instruction dependencies, with procedure slices being stored symbolically for precise interprocedural analysis. *Theorems 1 and 2 theoretically state that the accuracy of slicing methods based on SymPas and PDG/SDG (or SDG-IFDS) is essentially the same, although the presentation of the final slices are different.*

Table 4. **The comparison of static backward slice algorithms**

|  | Time Complexity | Space Complexity | Slice Computation | Dependency Structure | Summary Computation |
|---|---|---|---|---|---|
| **Weiser** [1] | $O(p \times n_{max} \times e_{max} \times c_{max} \times x^2)$ | $O(p^2 \times n_{max}^2)$ | *backward* dataflow | dataflow set | *NO* |
| **SDG-based** [9] | $O(p \times n_{PDG} \times e_{PDG} + c \times e_{PDG} \times x + c \times c_{max}^2 \times x^4)$ | $O(p^2 \times n_{PDG}^2)$ | graph reachability | PDG/SDG | Attribute grammar |
| **SDG-IFDS-based** [12] | $O(p \times n_{PDG} \times e_{PDG} + \boldsymbol{p \times e_{PDG} \times x} + c \times x^3)$ | $O(p^2 \times n_{PDG}^2)$ | graph reachability | PDG/SDG | IFDS |
| **SymPas** | $O(p \times n_{max} \times e_{max} \times (b_{max} + r) + c \times r \times n_{max} \times x^3)$ | $O(p^2 \times n_{max}^2)$ | *forward* dataflow | IDT | *Symbolic slice* |

About the efficiency of these static slicing methods, the total time cost of SDG-based *HRB algorithm* [9] is $O(p \times n_{PDG} \times e_{PDG} + c \times e_{PDG} \times x + c \times c_{max}^2 \times x^4)$, where $c_{max}$ is the largest number of call sites in any procedure, $c$ the number of call instructions, $x$ the number of global variables or formal parameters, $n_{PDG}$ and $e_{PDG}$ the largest number of nodes and edges in any procedure's PDG [2], respectively. Whereas, the SDG-IFDS-based *RHS algorithm* [10] takes time $O(p \times n_{PDG} \times e_{PDG} + p \times e_{PDG} \times x + c \times x^3)$, which is asymptotically faster than HRB algorithm (since generally $p < c$ ) [12]. It has one more term (i.e. $p \times e_{PDG} \times x$) than the time cost of our backward SymPas, as shown in Table 4. In addition, $e_{PDG} = O(n_{PDG}^2)$, $n_{PDG} = O(n_{max})$. So their space costs are asymptotically the same.

Although the extended IFDS algorithm in [13] constructs the exploded supergraph on demand, the reachability algorithms require the whole exploded supergraph to be constructed ahead of time. In other words, SDG-IFDS-based slicing exhaustively computes all nodes reachable along valid paths, rather than answering reachability queries for its reachable subgraph. In contrast, our symbolic slicing SymPas constructs the related tables (such as the procedure symbolic slice table $T$ and the instruction dependency table $L$) on demand, without requiring these whole tables to be constructed ahead of time, which will be shown in the experiments in the next section.

## 5.3 Empirical Evaluation

*5.3.1 Benchmarks.* To evaluate the efficiency of our symbolic slicer **SymPas** with two other IR slicers implemented above (i.e., **SDG-IFDS** and **MWeiser**), we automatically generated IR backward (*forward*) slicing criteria for single variable, to slice on the last (*first*) instruction of the main procedure for each global variable, and on the last (*first*) instruction of each procedure for each local variable allocated/declared in the procedure. We use fourteen benchmark programs. Five are from the Mälardalen WCET (Worst-Case Execution Time) benchmark [21] including *lms*, *compress*, *ndes*, *adpcm*, and *statemate*; two from SPECint 2000 benchmark including *181.mcf* and *256.bzip2*; five from real open-source programs including *time*-1.7, *gitview* from gnuit-4.9.5, *pkg-config*-0.26, *barcode*-0.99 and *byacc*; and two programs *loop50* and *loop100* are adapted from the while programs in [5] with 50 and 100 while-loop statements, respectively. The basic statistics of these benchmark programs are listed in Table 5, where LOC denotes the number of lines of front-end source code, $n$ the number of IR instructions compiled from front-end source code and, $b$, $p$ and $v$ have the same meaning as in Section 5.1. For other information such as their source codes (exclude SPECint), call graphs, SDG graphs, detailed forward/backward slice results, please see [20].

*5.3.2 Research Questions.* We want to evaluate the slicing results of our symbolic slicing tool (**SymPas**) to determine if accuracy slices are produced, and are produced efficiently. For the efficiency of these static slicing tools, here we compare our symbolic slicing method (SymPas) with three existing slicing methods mentioned in this paper: *Weiser*, *HRB* (SDG-based) and *RHS* (SDG-IFDS-based) algorithms. Since SDG-IFDS-based methods have been proven to be asymptotically faster than original SDG-based methods, we focus on evaluating slices of three slicer tools: SymPas, SDG-IFDS and MWeiser, by taking into consideration time and space efficiency, as shown in research questions *RQ1* and *RQ2* below.

For slicer accuracy, as mentioned before, SymPas is a context-sensitive slicing method via dataflow analysis on LLVM. It uses intuitive data structure IDT to represent instruction dependencies, with procedure slices being stored symbolically for precise interprocedural analysis. Theorem 1 and 2 theoretically state that the accuracy of slicing methods based on **SymPas** and PDG/SDG (or SDG-IFDS) is essentially the same, although the presentations of the final slices are different. In [30], Binkley et al observed that from 43 programs (ranging up to 136,000 LOC in size), *for the most precise slicer, the average slice contains just under one third of the program*. In order to check whether our slicing tool also satisfies this rule of precise slicer, we evaluate the slice size, i.e. the number of IR instructions (here possibly including parameter symbols), which is related to slice quality [30], as shown in research question *RQ3* below.

In our evaluation, we intend to answer three key research questions:

  1) *RQ1*: For backward slicing, is our **SymPas** more efficient than **SDG-IFDS** and **MWeiser** in terms of time and space costs?

  2) *RQ2*: For forward slicing, how about the efficiency of **SymPas** when compared with **SDG-IFDS**?

  3) *RQ3*: Is **SymPas** highly size-scalable?

*5.3.3 Experiment Results.* The performance statistics of these three static slicers is shown in Table 5, where programs are sorted by number of IR instructions (Column "*n*"), including results of backward and forward slicing. For each slicing algorithm, we report its running CPU time ("T" columns) and memory cost ("S" columns). From Table 5, we draw the following conclusions:

● For the Weiser algorithm, which only supports backward slicing, it always needs much more time and space to calculate the slice table of all single variables, since it requires repeated iterations for computing the slice of each variable, without reusing intermediate results as done in **SymPas** or **SDG-IFDS**. In addition, **MWeiser** analyses each procedure in an arbitrary order, whereas **SymPas** and **SDG-IFDS** traverse each SCC in the call graph bottom-up, which makes it easier to share intermediate values between analyses, thus only traversing the call graph once.

● For static backward slicing, on average, compared with SDG- and IFDS-based slicer **SDG-IFDS**, *our symbolic slicer SymPas reduced 84.0% in running time and 74.0% in space overhead.*

- For forward slicing, **SymPas** has slight advantage over **SDG-IFDS** on average, because it needs extra time cost $O(r \times p^2 \times n_{max}^2)$ to generate forward slice table from the IDT table in **SymPas**.

- For the scalability of slice size (size-scalability), we study SymPas slices from 66 programs (shown in [20]), ranging up to 336,859 instructions in size. The results (partly shown in Fig. 11) show that, *the average backward-SymPas slice contained 22.4% of the system, and 25.2% in forward SymPas. As expected, SymPas is highly size-scalable.*

In summary, for *RQ1*, we conclude that our symbolic backward slicing performs better than existing slicing methods such as SDG-based and Weiser slicing. For *RQ2*, **SymPas** has slight advantage over **SDG-IFDS** in forward slicing. For *RQ3*, **SymPas** has high slice-size scalability.

Table 5. **The efficiency evaluation of SymPas**
(LOC: #code-lines, *n*: #IR-instructions, *b*: #branches, *p*: #procedures,
*v*: #variables, T: time (sec.), S: space (MB), ∞: timeout (30 min.), —: N/A )

| System | LOC | n | b | p | v | Backward Slicing ||||||  Forward Slicing ||||
|---|---|---|---|---|---|---|---|---|---|---|---|---|---|---|---|
|  |  |  |  |  |  | SymPas || SDG-IFDS || MWeiser || SymPas || SDG-IFDS ||
|  |  |  |  |  |  | T | S | T | S | T | S | T | S | T | S |
| lms.c | 271 | 353 | 15 | 8 | 41 | 0.06 | 2 | 0.56 | 5 | 4.05 | 18 | 1.03 | 6 | 0.65 | 5 |
| compress.c | 521 | 494 | 36 | 9 | 57 | 0.43 | 8 | 5.67 | 59 | 52.28 | 72 | 3.70 | 43 | 6.20 | 59 |
| ndes.c | 238 | 586 | 26 | 5 | 62 | 0.19 | 5 | 1.44 | 22 | 16.51 | 63 | 1.00 | 20 | 1.90 | 21 |
| loop50.c | 364 | 819 | 50 | 1 | 6 | 10.52 | 59 | 273.6 | 2,064 | 322.1 | 2,858 | 90.74 | 530 | 262.26 | 2,010 |
| time | 1,395 | 953 | 55 | 6 | 66 | 0.64 | 15 | 18.51 | 286 | 144.3 | 374 | 5.16 | 143 | 17.99 | 285 |
| adpcm.c | 875 | 1,118 | 37 | 17 | 178 | 0.61 | 14 | 29.96 | 69 | 487.3 | 533 | 7.50 | 37 | 48.88 | 69 |
| statemate.c | 1,276 | 1,271 | 179 | 8 | 106 | 3.16 | 48 | 74.82 | 659 | ∞ | — | 37.91 | 492 | 72.03 | 661 |
| loop100.c | 714 | 1,619 | 100 | 1 | 6 | 105.2 | 361 | ∞ | — | ∞ | — | 1,310.36 | 4,023 | ∞ | — |
| 181.mcf | 4,820 | 2,916 | 179 | 26 | 186 | 2.56 | 30 | 24.47 | 74 | ∞ | — | 41.38 | 178 | 32.07 | 79 |
| gitview | 976 | 4,980 | 421 | 150 | 431 | 115.9 | 1,206 | 1,088.9 | 1,266 | ∞ | — | ∞ | — | 1,681.8 | 1,281 |
| 256.bzip2 | 8,014 | 6,258 | 479 | 74 | 421 | 59.37 | 272 | 972.0 | 2,085 | ∞ | — | 1,255.9 | 3,678 | 1,198.0 | 2,086 |
| pkg-config | 5,191 | 7,411 | 702 | 92 | 589 | 172.3 | 278 | 1,238.5 | 1,560 | ∞ | — | ∞ | — | 1,571.6 | 1,566 |
| barcode | 30,472 | 8,326 | 655 | 68 | 491 | 57.47 | 213 | 758.1 | 1,028 | ∞ | — | 161.71 | 1,154 | 912.86 | 1,025 |
| byacc | 23,906 | 13,953 | 1,141 | 208 | 875 | 304.2 | 280 | ∞ | — | ∞ | — | 1,658.6 | 1,640 | ∞ | — |
| Average | 5,645 | 3,647 | 291 | 48 | 25 | 59.49 | 199 | 373.8 | 765 | 171.1 | 653 | 381.25 | 995 | 483.87 | 762 |
| GeoMean | 1,717 | 1,981 | 132 | 16 | 11 | 6.68 | 56 | 57.34 | 248 | 65.53 | 190 | 38.07 | 239 | 69.07 | 248 |

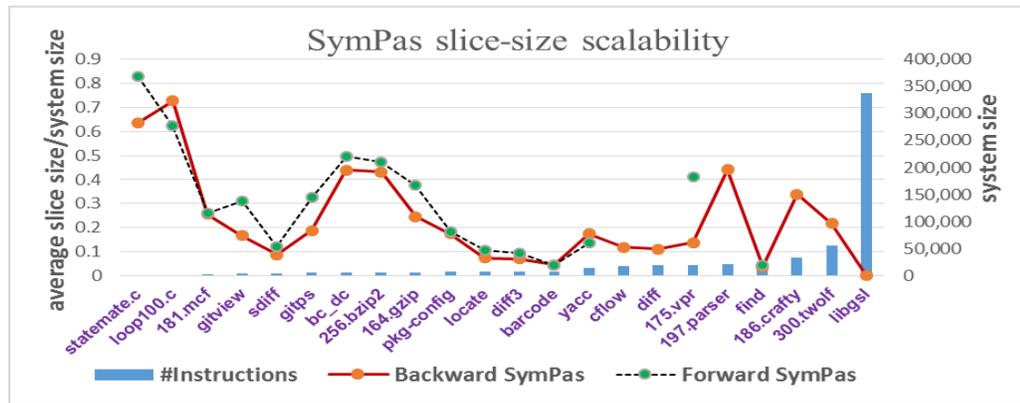

Fig. 11. **The high size scalability of SymPas.**

## 6 RELATED WORK

Weiser [1] first introduced program slicing. His interprocedural slicing algorithm is a simple extension of the intraprocedural one, and is easy implemented as it directly makes use of the intraprocedural slicing algorithm. But it cannot address the calling-context problem. Thus, it may produce imprecise slices.

Subsequently, Hwang et al. [31] proposed an iterative solution for interprocedural static slicing based on replacing recursive calls by instances of the procedure body. This approach does not suffer from the calling-context problem because expansion of recursive calls does not lead to considering infeasible execution paths. However, Reps [32][33] has shown that for a certain family of recursive programs, this algorithm takes exponential time in the length of the program.

In 1990, Horwitz et al. [9] proposed an interprocedural slicing algorithm based on the SDG representation. Their algorithm involves two steps: firstly, it constructs the corresponding SDG with summary edges, which represent transitive dependences due to procedure calls; secondly, it computes slices through two-phase traverses on the SDG. This slicing algorithm can address the calling-context problem, but the construction of SDG may be complex, as mentioned in Section 1. Subsequently, Lakhotia [34], and Livadas et al. [35][36] improved upon Horwitz's algorithm. However, the efficiency of the algorithm doesn't be improved essentially, and the complication of constructing SDG and the deficiency of language flexibility remain in these algorithms.

In 2012, Alomari et al. [37][38] presented a highly efficient, light-weight, forward static slicing approach, with no need to build the complex PDG/SDG but instead dependence and control information is computed as needed while computing the slice on a variable. Their slices are special forward slices, called *forward decomposition slices*, each with respect to a variable *v* is the union of the static forward slices at the set of statements that define *v*. These slices are executable, and how to extend their method to computing backward slices or non-executable slices is their future work. To avoid building PDGs/SDGs, we in [39]-[46] presented *monadic slicing*, which is the formal slicing method based on modular monadic semantics [47] of the program. It requires users to know esoteric techniques such as monads [48][49] and monad transformers [50], since we should give complete monadic semantic descriptions of a program before computing its slices. For the same purpose of avoiding the whole exploded dependence graphs, in this paper, we proposed *symbolic slicing*, where the intuitive data structure IDT (instruction dependency table) is used to represent instruction dependencies, and program slices are stored symbolically rather than procedure being re-analysed. We use symbolic slice of procedure parameters to propagate data dependence of parameters in a similar way as parameter passing, thus alleviating the calling-context problem.

In 2014, Binkley et al. [56] mentioned that *how to slice programs written in multiple languages* is still an open long-standing challenge. To solve this challenge for dynamic slicing, Binkley et al. proposed in [56] a language-independent slicing technique, *Observation-Based Slicing* (ORBS), by deleting program statements, executing the candidate slice with some inputs, and observing the behaviour for a given slicing criterion. ORBS can slice multi-language systems through leveraging the existing build tool-chain. Maybe this ORBS method can be extended to multi-language static slicing by the union of dynamic slices for many test cases, i.e. union slices [57] . Using the modern compilation framework LLVM, we attempt in this paper to address this challenge for static slicing, by presenting a new context-sensitive slicing technique, called *Symbolic Program Slicing*. We're slicing in a language-agnostic manner using LLVM IR, so one of the benefits is to being able to address Binkley's challenge for slicing multi-language systems.

In 2015, Lisper et al. [58] presented a light-weight interprocedural algorithm for backward static slicing based on a variant of the SLV (strongly live variables) analysis, which is an alternative data flow analysis for computing data dependencies. Their algorithm is context-sensitive, but only works for programs without (direct or indirect) recursive procedures. In addition, along with the progress of SLV slicing analysis, many new slicing criterions will be generated. In comparison, our SymPas does not have the limitation of working only for non-recursive procedures; it eliminates the need to maintain and generate many slicing criterions.

There are three existing slicing tools for LLVM IR: Dg [51], Giri [52][53] and LLVMSlicer [54][55]. Dg aims to generate a dependence graph for IR programs, which can be used to compute static IR slices via graph reachability. Dg is now under hard developement, esp. in collecting interprocedural summary information. Giri is actually a dynamic backward slicing compiler pass in LLVM. LLVMSlicer is based on *Weiser algorithm*. The LLVMSlicer code was written for the specific purpose (of checking properties described by finite state machines), with turning off of the IR optimizations, so it is not flexible enough to be used by others.

# 7 CONCLUSIONS

In this paper, we have proposed *Symbolic Program Slicing* (SymPas), which is a *light-weight context-sensitive slicing* method via forward dataflow analysis on LLVM. SymPas uses intuitive data structure IDT to represent instruction dependencies. It works interprocedurally by storing a form of symbolic slice for precise interprocedural analysis, thus alleviating the calling-context problem. Experiments show that SymPas is both efficient in running time and space usage, and highly scalable in slice size.

Furthermore, *Equation 1 in Section 2 is actually the key idea of SymPas*. In intraprocedural SymPas, Equations 2 and 3 specify the data-dependence part in Equation 1; Algorithm 1 is the pseudo-code implementation of Equation 1 for slicing single-procedural programs. In interprocedural SymPas, Equations 5 and 11 respectively specify control-dependence part and data-dependence part in Equation 1 at call sites; Algorithm 2 can be seen as the pseudo-code implementation of Equation 1 for slicing multi-procedural programs.

The symbolic slice can help to obtain accurate parameter-dependent information and the callees' influence on callers, so as to avoid the calling-context problem. A symbolic slice table (procedure slice summary) of a procedure summarizes the slice analysis of the procedure, which can be re-used at multiple call sites without re-analysing the whole procedure. The benefit of symbolic parameters in a symbolic slice is that instantiating is much faster than re-analysing the slice of a procedure. Thus this improves the flexibility and modularity of slicing.

The symbolic slicing algorithms are effectively demand-driven. By using a "lazy" computation strategy for symbolic slices (parameterised), only those symbolic slices directly relevant to the slicing criterion are performed. Instead of constructing the whole PDG/SDG or exploded supergraph ahead of time, our symbolic slicing analyses one procedure at a time in call-graph dependence order (starting at the leaf-procedures); and instead of repeatedly slicing a procedure called from multiple call sites, we calculate a procedure symbolic slice which can be just used when needed.

*SymPas Applications.* Program slicing has two main uses: 1) program comprehension, for which slicing at the source-code level is vital and 2) program analysis and optimization as part of a compiler. We address the latter using LLVM as the vehicle, i.e. program slicing of LLVM IR. In fact, IR slice results could be easily applied to generate the slices of source code, by extracting source codes from sliced IRs with line number information in the metadata of each IR instruction.

SymPas has the following key applications:

1) SymPas can be used to generate the slices of LLVM front-end source languages such as C (shown in [20]), by extracting source codes from sliced IRs with line number information in MetaData of each IR instruction;

2) SymPas can help *statically slicing multi-language systems, which is still an open long-standing challenge for program slicing* [56], by combining (with the *llvm-link* tool) their IR instructions generated from existing compiler tools such as Clang [59] for C/C++, VMKit [60] for Java, Numba [61] for Python, and JXcore [62] for JavaScript;

3) SymPas can be used for IR optimizations as an analysis pass in the LLVM framework; and

4) With SymPas, we can measure and evaluate LLVM projects using program slicing metrics [63].

At present, software systems keep growing in size and complexity, and get increasingly hard to understand and maintain, so software module metrics are very important in software development and maintenance. We can use code metrics as a way to quantitatively measure software qualities, thereby helping us better understand and maintain software systems. There are many code metrics related with the design or source code properties of programs, such as size, complexity, coupling, cohesion. To capture more fine-grained program properties, program slicing information was used in program code metrics, called *program slicing metrics*. The program slicing metrics measure the size, complexity, coupling, and cohesion properties of programs based on program slices. Based on IR instruction slices from SymPas, we may obtain more fine-grained code metrics than those from front-end source program slicers.

*SymPas Limitations.* The limitations of SymPas includes:

1) Fixed slice criterions, <$p$, $v$>, where $v$ is a single variable of interest, $p$ the program end point for backward slicing (or the program start point for forward slicing), and

2) Restriction on LLVM front-end compilers. For example, the C fragment "a=3; b=a+2" will be compiled by LLVM Clang to the IR instruction "store i32 5, i32* %b", resulting in no dependency of b and a.

*Future Work.* The future work of SymPas includes:

1) More precise symbolic slices by using more advanced alias analyses,

2) More effective data structures such as OBDD (Ordered Binary Decision Diagram) [22] for related tables (e.g. *L* and *S*),
3) Comparisons with existing C slicers such as CodeSurfer [64] in terms of power and scalability,
4) Dynamic symbolic slicing algorithms, and
5) Extension SymPas to support slicing object-oriented, distributed and concurrent programs.

## ACKNOWLEDGMENTS

This work was partially supported by the National Natural Science Foundation of China (No. 61300054) and the Qing Lan Project of Jiangsu Province. We would like to show our gratitude to Prof. Alan Mycroft of Cambridge University and Dr. Hareton Leung of Hong Kong Polytechnic University, for their helpful and constructive comments that greatly improved the manuscript.

## APPENDIX: PROOF OF THEOREMS

From the PDG-based definition of program slicing, the *PDG* for a program *P*, denoted by *G*, is a directed graph whose vertices represent the assignment statements and control predicates that occur in program *P*, and whose edges represent dependences (control dependences or data dependences) between program components. A *control dependence* from predicate *w* to a program component *v*, denoted $w \rightarrow_c v$, means that during execution, whenever *w* is evaluated and its value matches *v*, then *v* will be executed. A *data dependence* from a program component $v_1$ to another program component $v_2$, denoted $v_1 \rightarrow_f v_2$, means that the program's computation might be changed if the relative order of the components $v_1$ and $v_2$ were reversed. The *SDG*, which is an extension of PDGs, represents programs in a language that includes procedures and procedure calls. A SDG incudes a main PDG which refers to the system's main program, and other PDGs which refers to the system's auxiliary procedures and some additional edges such as call (control) edges, parameter-in edges, parameter-out edges and summary edges.

For a vertex *i* of a PDG *G*, the slice of *G* with respect to *i*, written as *G*/*i*, is a graph containing all vertices that can reach *i* via flow or control edges:
$$V(G/i) = \{w \mid w \rightarrow^*_{c,f} i \wedge w \in V(G)\}.$$
The edges in *G*/*i* are essentially those in the subgraph of *G* induced by *V*(*G*/*i*). We have
$$E(G/i) = \{(v \rightarrow_f w) \mid (v \rightarrow_f w) \in E(G) \wedge v, w \in V(G/i)\} \cup \{(v \rightarrow_c w) \mid (v \rightarrow_c w) \in E(G) \wedge v, w \in V(G/i)\}$$
where *V*(*G*) and *E*(*G*) are respectively a set of vertices and edges of *G*; $w \rightarrow_c v$ and $w \rightarrow_f v$ denote respectively a control dependence and flow (data) dependence from vertex *w* to vertex *v*.

For $w \rightarrow_c v$, the source of a control dependence edge, e.g. *w*, is always the entry vertex or a predicate (branch) vertex. For $w \rightarrow_f v$, there is a path in the standard control-flow graph for the program by which the definition of *x* at *w* reaches the use of *x* at *v*.

In our symbolic slicing algorithms (Algorithm 1 and 2), once an instruction is analyzed, its corresponding intermediate label set $L^k(w)$ will be computed followed by Equation 1. Moreover, the size of (i.e., the number of instructions in) intermediate set $L^k(w)$ at the same vertex will increase with the number of times (*k*) of its computation, i.e., $L^0(w) \subseteq L^1(w) \subseteq L^2(w) \subseteq \ldots \subseteq L^k(w) \subseteq \ldots \subseteq L(w)$, where *L*(*w*) denotes the final backward dependences of *w* after analyzing its corresponding program.

For two reachable vertices *w* and *v* (i.e., $w \rightarrow^*_{c,f} v$), there exists an executable path on which, whenever the program component represented by *w* is analyzed, the one represented by *v* will be analyzed (although perhaps not immediately).

Below, we use instruction vertices/nodes in PDG to denote the vertices corresponding to a LLVM IR instruction, excluding *entry vertex*, *initial definition vertex*, and *final use vertex*.

LEMMA 1. For two instruction vertices, *w* and *v*, in a PDG, if there is a direct edge from *w* to *v*, i.e., $w \rightarrow_{c,f} v$, then we have $L^k(w) \subseteq L^{k+1}(v)$ for some *k*.

PROOF. There are two cases to consider:

*Case 1*. If the vertex *w* represents a branch statement, i.e., $w \rightarrow_c v$, then $w \in CD(v)$, $L^k(w)$ will pass into $L^{k+1}(v)$ at the vertex *v* through the term $\bigcup_{j \in CD(v)} L^k(j)$ in Equation 1. This means $L^k(w) \subseteq L^{k+1}(v)$.

*Case 2*. If the vertex $w$ represents the statement that assigns the variable $x$ used in the vertex $v$, i.e. $w \to_f v$, $x \in \text{DEF}(w)$, and $x \in \text{REF}(v)$, then from Equation 2, $S^k(x) = L^k(w)$, where $S^k(x)$ is the corresponding slice of $x$ at the vertex $w$. Thus, the term $\cup_{x \in \text{REF}(v)} S^k(x)$ in Equation 1 will make $L^{k+1}(v)$ include $L^k(w)$, i.e. $L^k(w) \subseteq L^{k+1}(v)$. ∎

LEMMA 2. *For two instruction vertices, $w$ and $v$, in a PDG, if $w$ can reach $v$, i.e., $w \to^*_{c,f} v$, then we have $L^k(w) \subseteq L(v)$ for some $k$.*

PROOF. There are two cases to consider:

*Case 1*. $w \to_{c,f} v$, i.e., vertex $w$ reaches directly vertex $v$ via a flow or control edge. In this case, by Lemma 1, $L^k(w) \subseteq L^{k+1}(v)$ for some $k$, then $L^k(w) \subseteq L^{k+1}(v) \subseteq L(v)$, i.e. $L^k(w) \subseteq L(v)$ for some $k$.

*Case 2*. $w \to^+_{c,f} v$, i.e., vertex $w$ reaches indirectly vertex $v$ via two or more flow or control edges. If $w$ reaches $v$ through vertex $u$, i.e., $w \to_{c,f} u \to_{c,f} v$, then $L^k(w) \subseteq L^{k+1}(u)$ and $L^{k+m}(u) \subseteq L^{k+m+1}(v)$ by Lemma 1. Therefore, if $w \to_{c,f} u \to_{c,f} v$, then $L^k(w) \subseteq L^{k+m+1}(v) \subseteq L(v)$. Similarly, we can have $L^k(w) \subseteq L(v)$ in the case of $w \to^+_{c,f} v$. ∎

THEOREM 1. **Two PDG vertices (excluding *entry* vertices), $w$ and $v$, are reachable, i.e. $w \to^*_{c,f} v$, iff (if, and only if), $w \in L(v)$.**

PROOF. (1) "Only If" implication.

By Lemma 2, if $w \to^*_{c,f} v$, then $L^k(w) \subseteq L(v)$ for some $k$. Since $w \in L^0(w) \subseteq \ldots \subseteq L^k(w)$, we have $w \in L^k(w) \subseteq L(v)$, i.e. $w \in L(v)$.

(2) "If" implication.

From Equation 1, if vertex $w$ is in the final backward dependence result of vertex $v$, i.e. $w \in L(v)$, we have either $w \in \cup_{j \in \text{CD}(v)} L^k(j)$ or $w \in \cup_{x \in \text{REF}(v)} S^k(x)$ for some $k$. If $w \in \cup_{j \in \text{CD}(v)} L^k(j)$, then there exists vertex $v_1$ such that $v_1 \to_c v$ and $w \in L(v_1)$. If $w \in \cup_{x \in \text{REF}(v)} S^k(x)$, according to Equation 2, there exists vertex $w_1$, which represents an assignment instruction to variable $x_1$, i.e. $x_1 \in \text{DEF}(w_1)$, such that $w_1 \to_f v$, $x_1 \in \text{REF}(v)$, $S(x_1) = L(w_1)$, and $w \in L(w_1)$. Similarly, for the vertex $v_1$ (or $w_1$), there exists $v_2$ such that $v_2 \to_{c,f} v_1$ and $w \in L(v_2)$, (or exists $w_2$ such that $w_2 \to_{c,f} w_1$ and $w \in L(w_2)$). Go on in this way until $v_n = w$ and $v_n \to_{c,f} v_{n-1} \to_{c,f} v_{n-2} \to_{c,f} \ldots \to_{c,f} v_1 \to_c v$, (or $w_n = w$ and $w_n \to_{c,f} w_{n-1} \to_{c,f} w_{n-2} \to_{c,f} \ldots \to_{c,f} w_1 \to_f v$). Thus, $w \to^*_{c,f} v$, that is, vertex $w$ can reach vertex $v$.

So there exists control or data dependence (directly or indirectly) from vertex $w$ to vertex $v$, i.e. $w \to^*_{c,f} v$. ∎

From Theorem 1, we have two corollaries as follows.

COROLLARY 1. *For two PDG vertices (instructions), $w$ and $v$, there is a direct (control or data dependence) edge from $w$ to $v$, i.e. $w \to_{c,f} v$, iff (if and only if) $L^k(w) \subseteq L^{k+1}(v)$ for some $k$.*

PROOF. In fact, Lemma 1 is the 'Only If' implication of this lemma. So we just need to proof its 'If' implication as follows.

If $L^k(w) \subseteq L^{k+1}(v)$ for some $k$, then $w \in L^k(w) \subseteq L^{k+1}(v) \subseteq L(v)$, i.e. $w \in L(v)$. From Theorem 1, we have $w \to^*_{c,f} v$. According to the meaning of $k$, which corresponds to the level of indirect effect, we know that the value of indirect-effect level from $w$ to $v$ is 1, i.e. $w \to_{c,f} v$. ∎

COROLLARY 2. *The accuracy of the intra-procedural symbolic slicing algorithm (Algorithm 1) is same with that of PDG-based slicing algorithms.*

PROOF. By Theorem 1 ('Only If'), as for two reachable vertices $w$ and $v$, i.e., $w \to^*_{c,f} v$, we have $w \in L(v)$. Thus, if the program component represented by $w$ goes into the final slice for a variable in program component represented by $v$ because of the reachability from $w$ to $v$, then our symbolic slicing method can capture $w$ in the corresponding slice of that variable at $v$. In other words, the dependences in a PDG can be captured, if need be, by our symbolic slicing algorithm.

In other hand, Theorem 1 ('If') shows that if there exists vertex $w$ in $L(v)$ at vertex $v$ ($v \neq w$) by our symbolic slicing, then $w$ can reach vertex $v$ in their corresponding PDG. It means that no redundant dependences will be included in the slice result of symbolic slicing algorithm.

In short, our symbolic slicing method not only accurately captures all reachability information related, but also doesn't includes redundant reachability information. In this way, we can say that our symbolic slices are consistent with the PDG-based definition of program slicing. ∎

LEMMA 3. *In a SDG, for a *call-site* vertex $u$ in procedure $Q$, which calls procedure $P$, let $w$ be an *actual-in* vertex at $u$, with corresponding to formal parameter $x$ of $P$; let $v$ be an *actual-out* vertex at $u$, with corresponding to formal parameter $y$ of $P$. Then, there is a *summary* edge from $w$ to $v$, noted as $w \to_{su} v$, iff $x \in \text{SUMM}_P(y)$.*

PROOF. (1) "Only If" implication.

If there exists a *summary* edge from *actual-in* vertex $w$ to *actual-out* vertex $v$ at call site $u$ to $P$, i.e. $w \rightarrow_{su} v$, then the value of the corresponding variable in $v$ after the call to $P$ depends on the value of the corresponding variable in $w$ before the call to $P$, i.e. $v$ transitively depends on $w$. It means that the *formal-out* vertex of formal parameter $y$ depends on the *formal-in* vertex of formal parameter $x$ in $P$, so $x$ is in the backward slice of $y$ at the exit of $P$. According to our symbolic algorithm, the symbolic parameter, $l_x$, used to initialize the backward slice of $x$ at the entry of $P$ will be in the procedure symbolic slice of $y$, $T_P(y)$, at the exit of $P$, i.e. $l_x \in T_P(y)$. Thus, $x \in \text{SUMM}_P(y)$ by the definition of SUMM in Equation 8.

(2) "If" implication.

Similar as above analysis, if $x \in \text{SUMM}_P(y)$, then $l_x \in T_P(y)$. So the *formal-out* vertex of $y$ depends on the *formal-in* vertex of $x$ in $P$, resulting in the transitive flow dependence (*summary* edge) from *actual-in* vertex $w$ to *actual-out* vertex $v$ in $Q$, i.e. $w \rightarrow_{su} v$. ∎

LEMMA 4. *For two SDG vertices, $w$ and $v$, if there is a direct edge from $w$ to $v$, i.e., $w \rightarrow v$, then we have $L^k(w) \subseteq L^{k+1}(v)$ for some $k$.*

PROOF. By Lemma 1, we know that if $w \rightarrow_{c,f} v$, then $L^k(w) \subseteq L^{k+1}(v)$ for some $k$. So we need to show that if $w$ can directly reach $v$ by following a *call* edge, *parameter-in* edge, *parameter-out* edge or *summary* edge, respectively noted as $w \rightarrow_{ca} v$, $w \rightarrow_{pi} v$, $w \rightarrow_{po} v$, or $w \rightarrow_{su} v$, then $L^k(w) \subseteq L^{k+1}(v)$ for some $k$.

1) If $w \rightarrow_{ca} v$, then $w$ is a *call-site* vertex, and $v$ is the corresponding procedure *entry* vertex. Since *call* edges are a new kind of control dependence edge, $L^{k+1}(v) = L^k(v) \cup L^k(w)$ by Equation 1, thus $L^k(w) \subseteq L^{k+1}(v)$.

2) If $w \rightarrow_{pi} v$, then $w$ is an *actual-in* vertex of actual parameter $x$ at *call-site* vertex $u$, and $v$ is the corresponding *formal-in* vertex of formal parameter $y$ in called procedure $Q$. By Equations 1, 9 to 11, $L^{k+1}(v) = L^k(v) \cup L^k(u) \cup \text{SPEC}_u^k(l_y) = L^k(v) \cup L^k(u) \cup \text{U}_{z \in \text{REF}(x)} S^k(z)$; $L^k(w) = L^k(u) \cup \text{U}_{z \in \text{REF}(x)} S^k(z)$, so $L^k(w) \subseteq L^{k+1}(v)$.

3) If $w \rightarrow_{po} v$, then $w$ is a *formal-out* vertex of formal parameter $y$ in called procedure $Q$, and $v$ is the corresponding *actual-out* vertex of actual parameter $x$ at *call-site* vertex $u$. By Equations 1, 11 and 9, $L^{k+1}(v) = L^k(v) \cup L^k(u) \cup S^{k+1}(x) = L^k(v) \cup L^k(u) \cup T'_{Q,u}(y)$; $L^k(w) = L^k(u) \cup T'_{Q,u}(y)$, thus $L^k(w) \subseteq L^{k+1}(v)$.

4) If $w \rightarrow_{su} v$, then $w$ is an *actual-in* vertex of actual parameter $x$ at *call-site* vertex $u$ for calling $Q$, and $v$ is an *actual-out* vertex of actual parameter $y$ at $u$. Let $x'$ and $y'$ be $Q$'s corresponding formal parameters of $x$ and $y$, respectively. By Equations 1, 11 and 9, $L^{k+1}(v) = L^k(v) \cup L^k(u) \cup S^{k+1}(y) = L^k(v) \cup L^k(u) \cup T'_{Q,u}(y')$; $L^k(w) = L^k(u) \cup \text{U}_{z \in \text{REF}(x)} S^k(z)$. According to Lemma 3, we have $x' \in \text{SUMM}_Q(y')$. By Equations 9 and 10, $\text{U}_{z \in \text{REF}(x)} S^k(z) = \text{SPEC}_u^k(l_x) \subseteq T'_{Q,u}(y')$. In other words, the slice of each variable referenced in $x$ may pass into the slice result of $y$ through SPEC in Equation 10, $T'$ in Equation 9 and the third case in Equation 11. Then, $L^k(w) \subseteq L^{k+1}(v)$ for some $k$. ∎

LEMMA 5. *For two instruction vertices, $w$ and $v$, in a SDG, if $w$ can reach $v$ in procedure $Q$ by following control edges, flow edges and summary edges, i.e. $w \rightarrow^*_{c,f,su} v$, then $w \in L(v)$ and $L^k(w) \subseteq L(v)$ for some $k$.*

PROOF. By Lemma 2 and Theorem 1, we know that if $w \rightarrow^*_{c,f} v$, then $L^k(w) \subseteq L(v)$ and $w \in L(v)$. So we just need to show that if $w$ can reach $v$ by following some control or flow edges with one or more *summary* edges, i.e. $w \rightarrow^+_{c,f,su} v$, then $L^k(w) \subseteq L(v)$ and $w \in L(v)$.

Suppose $w$ reaches $v$ through *actual-in* vertex $u_1$ (of actual parameter $x$) and *actual-out* vertex $u_2$ (of actual parameter $y$) at *call-site* vertex $u$ in $Q$, such that $w \rightarrow^*_{c,f} u_1$, $u_1 \rightarrow_{su} u_2$ and $u_2 \rightarrow^*_{c,f} v$. According to Theorem 1 and Lemmas 1, 2 and 4, we have: $L^k(w) \subseteq L^{k+m}(u_1) \subseteq L^{k+m+1}(u_2) \subseteq L^{k+m+1+n}(v) \subseteq L(v)$, so $L^k(w) \subseteq L(v)$ and $w \in L(v)$. Similarly, we can have $L^k(w) \subseteq L(v)$ and $w \in L(v)$ in the case of $w \rightarrow^+_{c,f,su} v$. ∎

THEOREM 2. **For two instruction vertices, $w$ and $v$, in a SDG $G$, let $G/v$ be the backward slice of $v$ via SDG-based slicing algorithms. Then, $w \in G/v$ iff $w \in L(v)$.**

PROOF. (1) "Only If" implication.

According to the SDG-based algorithm, $G/v$ can be determined by traversing the SDG graph in two phases (started at vertex $v$): *Phase 1* and *Phase 2*. *Phase 1* determines all SDG vertices from which $v$ can be reached without descending into called procedures (via *parameter-out* edges). The *summary* edges (transitive flow dependences) guarantee that called procedures can be sidestepped, without descending into them. In *Phase 2*, the algorithm descends into all called procedures sidestepped in *Phase 1*, without ascending to call sites (via *call* edges or *parameter-in* edges), and determines the remaining vertices in the slice $G/v$. So for $w \in G/v$, there are two cases as follows to consider. Here we suppose that vertex $v$ is in procedure $P$.

*Case 1.* $w$ is included in the slice $G/v$ during *Phase 1*, i.e. $w$ can reach $v$ by *Phase 1*, noted as $w \to^*_{P1} v$.

*Phase 1* identifies those vertices that can reach $v$ by following *flow* edges, *control* edges, *call* edges, *parameter-in* edges and *summary* edges, and that are either in $P$ itself or in a procedure (say $Q$) that calls $P$ (either directly or transitively). So in this case, we know that vertex $w$ is either in $P$ or in $Q$. If vertex $w$ is also in $P$ and can reach $v$ by following *control* edges, *flow* edges and *summary* edges, i.e. $w \to^*_{c,f,su} v$, then according to Lemma 5, we have $w \in L(v)$.

If vertex $w$ is in a caller $Q$ that calls $P$ and $w \to^*_{P1} v$, then there exists *call-site* vertex $u$ in $Q$ for calling $P$, such that $w \to^*_{c,f,su} u$ and $u \to^*_{P1} v$. There are two cases for $u \to^*_{P1} v$:

1) $u$ reach *entry* vertex $w$ of $P$ by a *call* edge, i.e. $u \to_{ca} w$, and $w \to^*_{c,f} v$. By Theorem 1, Lemmas 4 and 5, $w \in L^k(u) \subseteq L^{k+1}(w) \subseteq L^{k+m+1}(v) \subseteq L(v)$, i.e. $\boldsymbol{w \in L(v)}$.

2) There exists *actual-in* vertex $w_1$ (of actual parameter $x$) at $u$, which reaches corresponding *formal-in* vertex $w_2$ (of formal parameter $y$) in $P$ by a *parameter-in* edge, i.e. $w_1 \to_{pi} w_2$, such that $u \to_c w_1$, $w_2 \to^*_{c,f,su} v$. By Theorem 1, Lemmas 4 and 5, $w \in L^k(u) \subseteq L^{k+1}(w_1) \subseteq L^{k+2}(w_2) \subseteq L^{k+m+2}(v) \subseteq L(v)$, i.e. $\boldsymbol{w \in L(v)}$.

Similarly, if vertex $w$ is in a caller $Q$ that transitively calls $P$, and $w \to^*_{P1} v$, we have $\boldsymbol{w \in L(v)}$.

Thus, if $w \in G/v$ because $w \to^*_{P1} v$, then we have $w \in L(v)$.

*Case 2.* $w$ is included in the slice $G/v$ during *Phase 2*, i.e. $w$ can reach $v$ by *Phase 2*, noted as $w \to^*_{P2} v$.

*Phase 2* identifies those vertices that can reach $v$ by following *flow* edges, *control* edges, *parameter-out* edges and *summary* edges, from a procedure $R$ (transitively) called by $P$ or from a procedure $R'$ called by procedures that (transitively) call $P$. So in this case, we know that vertex $w$ is either in $R$ or in $R'$. If vertex $w$ is in $R$ called by $P$ and $w \to^*_{P2} v$, then there exists *actual-out* vertex $w_1$ in $P$, which is reached from corresponding *formal-out* vertex $w_2$ in $R$ by a *parameter-out* edge, i.e. $w_2 \to_{po} w_1$, such that $w \to^*_{c,f,su} w_2$, $w_1 \to^*_{c,f,su} v$. By Theorem 1, Lemmas 4 and 5, $w \in L^k(w_2) \subseteq L^{k+1}(w_1) \subseteq L^{k+1+m}(v) \subseteq L(v)$, i.e. $\boldsymbol{w \in L(v)}$. Similarly, if vertex $w$ is in a callee $R$ called by $P$ transitively, and $w \to^*_{P2} v$, we have $\boldsymbol{w \in L(v)}$.

If vertex $w$ is in $R'$ called by procedure $Q$ that (transitively) calls $P$ and $w \to^*_{P2} v$, then there exists *actual-out* vertex $w_1$ in $Q$, which is reached from corresponding *formal-out* vertex $w_2$ in $R'$ by a *parameter-out* edge, i.e. $w_2 \to_{po} w_1$, such that $w \to^*_{c,f,su} w_2$, $w_1 \to^*_{P1} v$. By *Case 1* and Theorem 1, Lemmas 4 and 5, $w \in L^k(w_2) \subseteq L^{k+1}(w_1) \subseteq L^{k+1+m}(v) \subseteq L(v)$, i.e. $\boldsymbol{w \in L(v)}$.

Thus, if $w \in G/v$ because $w \to^*_{P2} v$, then we have $w \in L(v)$.

(2) "If" implication.

From Equations 1, 5 and 11, if vertex $w$ is in the final backward dependence result of vertex $v$, i.e. $w \in L(v)$, we have either $w \in \bigcup_{j \in CD(v)} L^k(j)$ or $w \in \bigcup_{x \in REF(v)} S^{k+1}(x)$ for some $k$. If $w \in \bigcup_{j \in CD(v)} L^k(j)$, then there exists vertex $v_1$ such that $w \in L^k(v_1)$, $v_1 \to_c v$, or $v_1 \to_{ca} v_2 \to_c v$, where $v_2$ is *entry* vertex of $P$. Thus, if $w \in L(v)$ because $w \in \bigcup_{j \in CD(v)} L^k(j)$ for some $k$, then $w \in L^k(v_1)$ and $v_1 \to^*_{P1} v$ for some $v_1$.

If $w \in \bigcup_{x \in REF(v)} S^{k+1}(x)$, according to Equation 11, there are three cases to consider: 1) there exists vertex $w_1$, which represents an assignment instruction to variable $x_1$, i.e. $x_1 \in DEF(w_1)$, such that $x_1 \in REF(v)$, $w \in S(x_1) = L(w_1)$, $w_1 \to_f v$, thus $w_1 \to^*_{P1} v$. 2) there exists *actual-in* vertex $w_1$ of actual parameter $y$ at *call-site* vertex $u$ for calling $P$, with corresponding to *formal-in* vertex $w_2$ of formal parameter $y'$, such that $w_1 \to_{pi} w_2$, $y' \in SUMM_P(x)$, $x \in REF(v)$, $w_2 \to^*_{c,f,su} v$, $w \in L(w_1)$ and $w \in \bigcup_{z \in REF(y)} S^k(z) = SPEC_u^k(l_{y'}) \subseteq T'_{P,u}(x) = S^{k+1}(x)$, thus $w_1 \to^*_{P1} v$. 3) there exists *actual-out* vertex $w_2$ of actual parameter $x$ at *call-site* vertex $u$ to call procedure $Q$, with corresponding to *formal-out* vertex $w_1$ of formal parameter $x'$, such that $w_1 \to_{po} w_2$, $w_2 \to^*_{c,f,su} v$, $x \in REF(v)$, $x' \in OUT(Q)$, $w \in L(w_1)$ and $w \in T_Q(x') \subseteq T'_{Q,u}(x') = S^{k+1}(x)$, thus $w_1 \to^*_{P2} v$.

Similarly, for above vertex $v_1$ (or $w_1$), there exists $v_2$ such that $v_2 \to^*_{P1} v_1$ and $w \in L(v_2)$, (or exists $w_2$ such that $w_2 \to^*_{P1,P2} w_1$ and $w \in L(w_2)$). Go on in this way until $v_n = w$ and $v_n \to^*_{P1} v_{n-1} \to^*_{P1} \ldots \to^*_{P1} v_1 \to^*_{P1} v$, (or $w_n = w$ and $w_n \to^*_{P1,P2} w_{n-1} \to^*_{P1,P2} \ldots \to^*_{P1,P2} w_1 \to^*_{P1,P2} v$). Thus, if $w \in L(v)$ because $w \in \bigcup_{x \in REF(v)} S^{k+1}(x)$ for some $k$, then $w \to^*_{P1,P2} v$, i.e. $w \in G/v$.

Therefore, if $w \in L(v)$, then $w \in G/v$. ∎